\documentclass[journal]{IEEEtran}
\usepackage{amsthm, amssymb, amsfonts,algorithm, algpseudocode, array}
\usepackage{pifont}
\usepackage[cmex10]{amsmath}
\usepackage{cancel,cite,color,comment,dblfloatfix}
\usepackage{enumerate,esint}
\usepackage{float}
\usepackage[T1]{fontenc}
\usepackage{graphicx}
\usepackage[latin9]{inputenc}
\usepackage{multirow,multicol}
\usepackage{tikz,tgtermes}
\usepackage[labelformat=simple]{subcaption}
\usepackage{pgfplots}
\usepackage{bbm}
\usepackage{dsfont}
\usepackage{hyperref}

\usetikzlibrary{backgrounds}
\usetikzlibrary{shapes, snakes}
\pgfplotsset{compat=1.11}
\allowdisplaybreaks
\newcommand{\dS}{{d^{(c, \epsilon)}}}
\newcommand{\dSh}{{\bar{d}^{(c, \epsilon)}}}

\newcommand{\R}{\mathbb{R}}

\newcommand{\W}{{\mathcal{W}}}
\newcommand{\hW}{{\overline{\mathcal{W}}}}

\theoremstyle{plain}
\newtheorem{thm}{Theorem}
\newtheorem{lem}[thm]{Lemma}
\newtheorem{prop}[thm]{Proposition}

\theoremstyle{definition}
\newtheorem{defn}{Definition}

\theoremstyle{remark}

\title{Graph GOSPA metric: a metric to measure the discrepancy between graphs of different sizes}
\author{Jinhao Gu, \'Angel F. Garc\'ia-Fern\'andez, Robert E. Firth, Lennart Svensson \thanks{J. Gu and A. F. Garc\'ia-Fern\'andez are with Department of Electrical Engineering and Electronics, University of Liverpool, Liverpool L69 3GJ (emails: \{jinhgu, angel.garcia-fernandez\}@liverpool.ac.uk). A. F. Garc\'ia-Fern\'andez is also with the ARIES research centre, Universidad Antonio de Nebrija, Madrid, Spain. R. E. Firth is with the STFC Hartree Centre, WA4 4AD Daresbury, UK (email: robert.firth@stfc.ac.uk). L. Svensson is with the Department of Electrical Engineering, Chalmers University of Technology, SE-412 96 Gothenburg, Sweden. This work was supported by the EPSRC Centre for Doctoral Training in Distributed Algorithms EP/S023445/1.}}
\begin{document}
\maketitle
\begin{abstract}
This paper proposes a metric to measure the dissimilarity between graphs that may have a different number of nodes. The proposed metric extends the generalised optimal subpattern assignment (GOSPA) metric, which is a metric for sets, to graphs. The proposed graph GOSPA metric includes costs associated with node attribute errors for properly assigned nodes, missed and false nodes and edge mismatches between graphs. The computation of this metric is based on finding the optimal assignments between nodes in the two graphs, with the possibility of leaving some of the nodes unassigned. We also propose a lower bound for the metric, which is also a metric for graphs and is computable in polynomial time using linear programming. The metric is first derived for undirected unweighted graphs and it is then extended to directed and weighted graphs. The properties of the metric are demonstrated via simulated and empirical datasets.
\end{abstract}
\section{Introduction}

Graphs are a type of structured data that can be defined as collections of objects, where certain pairs of objects have a relationship of some kind represented by edges \cite{Trudeau_book93}. As a flexible and rich representation, graphs are prevalent in real-world applications in multiple research fields, such as social networks \cite{wu2020graph}, physical systems \cite{sanchez2018graph, battaglia2016interaction}, and chemical reactions \cite{kvasnivcka1991reaction}. Additionally, graphs can also be applied in machine learning \cite{gama2019} and signal processing tasks \cite{mateos2019, Ortega_book22, domingos2020graph}. 

It is important to be able to measure the discrepancy between graphs using a distance or a loss function \cite{shit2022relationformer}. One option to measure the discrepancy between graphs is to first represent each graph as a graphon (also known as a graph limit) \cite{lovasz2012large}, which is a bounded symmetric measurable function defined in $[0,1]\times[0,1]$. One can then use a metric in the graphon space, such as the $L^p$ metric or the cut metric, to measure the difference between graphs \cite{lovasz2012large}. However, this approach does not consider node attributes and the resulting metrics on the graphon space cannot be easily interpreted in terms of edge errors, or difference in the number of nodes. In addition, there are many design variables to convert a graph into a graphon. 

Another option, used in \cite{lee2011computing}, to measure the difference between graphs of the same sizes is to obtain their single linkage matrices, and then use the Gromov-Haussdorf metric \cite{burago2001course}, also used in shape analysis \cite{memoli07}, to measure the difference between theses single linkage matrices. An alternative is to obtain the Laplacian matrices of the graphs and compute the Wasserstein distance of their induced Gaussian distributions, as in \cite{Maretic19}. This distance can measure structural properties of the graphs, but is limited to graphs with the same number of nodes, without node attributes.

A desirable characteristic of a distance function is that it operates directly on the space of interest (in our case graphs) and that it is a mathematically principled metric, as metrics are consistent with an intuitive notion of error. Metrics are functions that meet the properties of non-negativity, identity, symmetry and triangle inequality \cite{Apostol_book74}.
Distances that satisfy all the properties of a metric, but not the identity property, are commonly referred to as pseudometrics \cite{howes2012modern}. Pseudometrics play an important role in measuring the discrepancy between graphs that do not have node attributes. In this case, nodes are represented by arbitrary labels and these pseudometrics only consider the discrepancies in the edge connections so the identity property of metrics is not of relevance. Pseudometrics for graphs can nevertheless be used to define a metric for the equivalence class of isomorphic graphs \cite{howes2012modern}.

Graph distances can be divided into several types based on their properties: 1) distances for graphs with or without node attributes; 2) distances that are metrics for graphs of the same size (same number of nodes); 3) distances that are also metrics for graphs of different sizes; 4) distances that are pseudometrics for graphs of the same size; 5) distances that are pseudometric for graphs of different sizes. In the following, we review some commonly used graph distances.

A classic way to compute the dissimilarity between graphs is through graph edit distances (GEDs) \cite{sanfeliu1983, fischer2015approximation}. Based on costs for insertions, deletions, and substitutions of both nodes and edges, the GEDs compute the path of modifications to transform one graph into the other with minimum cost. A GED defines a metric if the individual cost functions for each edit operation are metrics \cite{Justice06}. GEDs are generally NP-hard to compute but there are approximate strategies to compute them in polynomial time, e.g., linear programming relaxation \cite{Justice06}, bipartite graph matching  \cite{riesen2009approximate}, and greedy algorithms  \cite{riesen2015greedy, saadah2020}. However, these approximations do not necessarily preserve the metric properties. 

It is also possible to define graph distances by computing the maximum common subgraph (MCS) between two graphs. For example, a GED can be defined based on the MCS \cite{bunke1997relation}. In addition, it is possible to define a pseudometric based on the MCS \cite{bunke1998graph}, in which distances between isomorphic graphs are zero. However, finding the MCS of two graphs is NP-hard, making it computationally challenging, especially for large graphs.

The chemical distance is a pseudo-metric that does not consider information on node attributes \cite{kvasnivcka1991reaction}. It is obtained by minimising the edge discrepancy between two graphs of the same size, whose mapping is represented via a permutation matrix  \cite{kvasnivcka1991reaction}. Finding this optimal mapping between two graphs is computationally expensive, and the computationally efficient approximations to estimate the optimal permutations \cite{El-Kebir15,lyzinski16} break the triangle inequality property. The Chartrand-Kubiki-Shultz (CKS) distance \cite{chartrand1998graph} for graphs of the same size consists of applying the chemical distance to the weighted graphs whose adjacency matrices contain pairwise shortest path distances between the nodes.

A family of graph distances for graphs of the same size that includes the chemical distance and the CKS distance was defined in \cite{Bento19}. We refer to these distances as generalised chemical distances (GCD). In its main form, the GCDs only consider edge mismatches and therefore define pseudometrics. The GCDs have the benefit that being computationally tractable. For example, an implementation based on the alternating direction method of multipliers (ADMM) for large graphs was proposed in \cite{Moharrer20}. Node attribute errors can be incorporated into the GCDs such that GCDs define metrics for graphs of the same size \cite{Bento19}. If graphs are of different sizes, we can apply the GCD by adding dummy nodes such that both graphs are of the same size, as done in the graph matching algorithm in \cite{Yan15} and in the integer programming GED metric in \cite{Justice06}. However, special care should be taken in how these dummy nodes are penalised to preserve the GCD metric properties. For example, in \cite{Moharrer20}, the dummy nodes have zero as an attribute and do not have edges with other nodes. This use of dummy nodes implies that the triangle inequality property is not met. For instance, if we compare two graphs without edges, we compare two sets of nodes, and defining a metric for sets of nodes requires more considerations \cite{Rahmathullah17}.

In this paper, we propose the graph generalised optimal sub-pattern assignment (GOSPA) metric to measure the discrepancy between graphs of different sizes with node attributes. The graph GOSPA metric is an extension of the GOSPA metric \cite{Rahmathullah17}, which is a metric for sets, to graphs. The graph GOSPA metric computes an optimal assignment between nodes by penalising attribute errors for assigned nodes and also penalising nodes that are left unassigned. Apart from penalties for nodes, the graph GOSPA metric includes penalties for edge mismatches in the graphs. As the graph GOSPA metric can leave nodes in both graphs without assignment, it does not require the introduction of dummy nodes to deal with graphs of different sizes. This is computationally beneficial as it lowers the dimensionality of the optimisation problem. In addition, it includes a hyperparameter $p$ to adapt the cost of outliers. We also propose a lower bound on the graph GOSPA metric that is also a metric and is computable in polynomial time using linear programming (LP). We first present the graph GOSPA metric for unweighted, undirected graphs, and then we extend it to weighted and also directed graphs. We also define an associated pseudometric for graphs that do not have node attributes.

The outline of the paper is as follows. In Section \ref{sec:problem_form}, we formulate the problem of designing a metric and provide the background. Section \ref{sec:graph_metric_def} presents the proposed metric based on optimal assignments. In Section \ref{sec:lp_graph_metric}, we present the LP metric and its decomposition in terms of localisation costs for properly assigned nodes, costs for missed nodes, false nodes, and edge mismatch. In Section \ref{sec:extesion to weighted and directed graphs}, we extend the proposed metric to directed and weighted graphs. In Section \ref{sec:experiments}, we analyse the proposed metric implementations via simulations and real datasets. Finally, conclusions are drawn in Section \ref{sec:conclusion}.

\section{Problem formulation and background}\label{sec:problem_form}
In this section, we formulate the problem of designing a metric for undirected and unweighted graphs and review the GOSPA metric.
\subsection{A metric for graphs} \label{sec:graph_metric}
An undirected, unweighted graph $(V,E)$ consists of a pair of sets: the vertex set $V$ and the edge set $E$. The vertex set $V=\left\{ x_{1},...,x_{n_X}\right\}$ is a set of vertices (also called nodes) with the $i$-th node denoted by $x_{i}\in\mathbb{X}$, where $\mathbb{X}$ is the single-node attribute space, which has an associated base metric $d(\cdot,\cdot)$. For example, if the attribute is a real vector, we can set $\mathbb{X}=\mathbb{R}^N$. Therefore, $V\in\mathcal{F}\left(\mathbb{X}\right)$, where $\mathcal{F}\left(\mathbb{X}\right)$ denotes the set of finite subsets of $\mathbb{X}$. Given the set of nodes, the edge set $E\left(V\right)\subseteq\mathbb{E}=\left\{ \left\{ x,y\right\} :x,y\in V,x\neq y\right\}$, where we consider graphs without self-edges. The space of these graphs is denoted by $\Omega$.

Our objective is to design a metric $d(\cdot, \cdot)$ to measure the difference between two graphs $X=(V_{X},E_{X})$ and $Y=(V_{Y},E_{Y})$. Without loss of generality, we assume that $X$ is the ground truth graph and $Y$ is the estimated graph, provided by some algorithm. Therefore, a metric enables us to rank different graph estimates according to their closeness w.r.t. the ground truth graph. 

Suppose $d(\cdot,\cdot)$ is a function such that $d : \Omega \times \Omega\rightarrow \mathbb{R}_+= [0,+\infty)$. The function $d(\cdot,\cdot)$ is a metric on $\Omega$ if it meets the following properties for all $X,Y,Z \in \Omega$ \cite{Apostol_book74}:
\begin{enumerate}
    \item (identity) $d(X,Y)=0$ if and only if $X=Y$;
    \item (symmetry) $d(X,Y)=d(Y,X)$;
    \item (triangle inequality) $d(X,Y) \leq d(X,Z) + d(Z,Y)$.
\end{enumerate}
The proposed metric will be based on the GOSPA metric \cite{Rahmathullah17}, which is a metric for sets, and we proceed to review it in the next subsection.

\subsection{GOSPA metric}\label{section:gospa_defn}
In this section, we review the GOSPA metric between two sets of nodes. Suppose that $V_X=\{x_1\hdots,x_{n_X}\}$ is the ground truth of set of nodes and $V_Y=\{y_1,\hdots,y_{n_Y}\}$ is the estimate of the corresponding set of nodes. 

We first define an assignment vector $\pi=\{\pi_{1},\hdots,\pi_{n_X} \}$ between sets $\{1,\hdots,n_X\}$ and $\{1,\hdots,n_Y\}$. The assignment vector meets $\pi_i \in \{0,\hdots, n_Y\}$ and if $\pi_{i}=\pi_{i'}=j>0$, then this implies that $i=i'$.  Here, $\pi_{i}=j$ implies that $x_{i}$ is assigned to $y_{j}$, and $\pi_{i}=0$ implies that $x_{i}$ is left unassigned. The set of all possible assignment vectors between $V_{X}$ and $V_{Y}$ is denoted $\Pi_{V_{X},V_{Y}}$.

\begin{defn}\label{gospa metric}
Given a base metric $d(\cdot,\cdot)$ in $\mathbb{X}$, a scalar $c>0$, and a scalar $p$ with $1 \leq p \leq \infty$, the GOSPA metric ($\alpha=2)$\cite{Rahmathullah17} between sets $V_X$ and $V_Y$ is 
\begin{equation}
\begin{aligned}
d^{\left(c\right)}\left(V_X,V_Y\right) 
 =&\min_{\pi\in\Pi_{V_X,V_Y}} \left(\sum_{i=1}^{n_{X}}d\left(x_{i},y_{\pi_{i}}\right)^{p} \right.\\
 & \left.+\frac{c^{p}}{2}\left(n_{X}+n_{Y}-2\sum_{i=1}^{n_{X}}\mathds{1}_{\pi_{i}>0}(\pi_{i})\right)\right)^{1/p}\label{eq:gospa metric}
\end{aligned}
\end{equation}
where $\mathds{1}_{A}(\cdot)$ is an indicator function of the set A. That is, $\mathds{1}_{A}(\cdot)=1$ if $x\in A$ or 0 otherwise.
\end{defn}

The first term in \eqref{gospa metric} represents the distance between node attributes for assigned nodes to the power of $p$. The second term represents the cost of unassigned nodes, to the power of $p$. Two assigned nodes $x_i$ and $y_{\pi_i}$ contribute to the overall error with $d(x_i, y_{\pi_i})^p$. An unassigned node in $V_X$ or $V_Y$ incurs a penalty of $c^p/2$. In multi-object tracking, assigned nodes would represent an object and its estimate with a localisation error $d(x_i, y_{\pi_i})$. An unassigned node in $V_X$ would represent a missed object, and an unassigned node in $V_Y$ would represent a false object.

The GOSPA metric is a metric for sets, and is not a metric for graphs as it does not take into account the corresponding set of edges. This implies that GOSPA for sets does not meet the identity property when we consider graphs. Nevertheless, the metric for graphs we propose will be based on extending the GOSPA metric to additionally consider the set of edges, so we refer to the proposed metric as graph GOSPA metric.
That is, the graph GOSPA metric will make an assignment between nodes in $V_X$ and $V_Y$ to penalise localisation errors for properly assigned nodes and the number of missed and false nodes. Additionally, the graph GOSPA metric will include an edge mismatch penalty depending on these assignments in a manner that the metric properties in Section \ref{sec:graph_metric} are preserved. 

\section{Graph metric based on optimal assignments}\label{sec:graph_metric_def}
This section is organised as follows. Section \ref{subsec:graph gospa defn} presents the graph GOSPA metric. Section \ref{subsec:examples} provides some examples to illustrate how the metric works.

\subsection{Graph GOSPA metric}\label{subsec:graph gospa defn}

The graphs $X=(V_{X},E_{X})$ and $Y=(V_{Y},E_{Y})$ have sets of nodes $V_{X}=\left\{ x_{1},...,x_{n_{X}}\right\} $ and
$V_{Y}=\left\{ y_{1},...,y_{n_{Y}}\right\} $. The corresponding adjacency
matrices to represent the sets of edges for the node orderings $x_{1:n_{X}},y_{1:n_{Y}}$
are $A_{X}$ and $A_{Y}$, respectively \cite{Ortega_book22}. We use $A_{X}(i,j)$ to represent the $(i,j)$ element of matrix $A_X$. Specifically $A_X(i,j)=A_X(j,i)=1$ if there is an edge between node $i$ and node $j$ of $X$, and $A_X(i,j)=A_X(j,i)=0$ if there is no edge.

\begin{defn}\label{defn:graph_gospa}
For $1 \leq p < \infty$, a scalar $c>0$, edge mismatch penalty $\epsilon>0$ and base metric $d(\cdot,\cdot)$ in the single node space $\mathbb{X}$, the graph GOSPA metric $d^{(c,\epsilon)}(X,Y)$ between two graphs $X$ and $Y$ is
\begin{flalign} \label{eq:graph_metric}
&d^{\left(c,\epsilon\right)}\left(X,Y\right) 
 =\min_{\pi\in\Pi_{V_X,V_Y}}\left(\sum_{i=1}^{n_{X}}d\left(x_{i},y_{\pi_{i}}\right)^{p} \right. \nonumber\\
 & \left.+\frac{c^{p}}{2}\left(n_{X}+n_{Y}-2\sum_{i=1}^{n_{X}}\mathds{1}_{\pi_{i}>0}(\pi_{i})\right)
 +e_{X,Y}\left(\pi\right)^{p}\right)^{1/p}
\end{flalign}
where the edge mismatch cost is given by
\begin{flalign} \label{eq:edge_cost}
    e_{X,Y}\left(\pi\right)^{p} & =\frac{\epsilon^{p}}{2}\sum_{i=1:\pi_{i}>0}^{n_{X}}\sum_{j=1:\pi_{j}>0}^{n_{X}}\left|A_{X}(i,j)-A_{Y}(\pi_{i},\pi_{j})\right| \nonumber\\
    & + \frac{\epsilon^{p}}{2}\sum_{i=1:\pi_{i}=0}^{n_{X}}\sum_{j=1:\pi_{j}>0}^{n_{X}} \left|A_X(i,j) \right| \nonumber\\ 
    & + \frac{\epsilon^{p}}{2}\sum_{i=1:\not\exists l:\pi_{l}=i}^{n_{Y}}\sum_{j=1:\exists l:\pi_{l}=j}^{n_{Y}}|A_{Y}(i,j)|.
    \end{flalign}

\end{defn}

We can see that, similarly to the GOSPA metric in \eqref{eq:gospa metric}, \eqref{eq:graph_metric} is an optimisation problem over all assignment vectors. The function to optimise is the same as in \eqref{eq:gospa metric}, including node attribute errors (localisation errors) and costs for missed and false nodes, but adding an extra penalty for the edge mismatch $e_{X,Y}(\pi)$.
There are two types of edge mismatch costs:
\begin{itemize}
    \item An edge cost created by two pairs of properly assigned nodes (two nodes in $X$ and two nodes in $Y$). If the edge exists in one graph but it does not exist in the other graph, this contributes to an edge mismatch penalty. The corresponding cost is $\epsilon^p$ and the sum over all these costs is given by the first line in \eqref{eq:edge_cost}.
    \item For both graphs, each edge connecting an assigned node and an unassigned node contributes to a half-edge mismatch penalty. The corresponding cost is $\epsilon^p/2$. The sum over all such costs for edges in $X$ is given by the second line of \eqref{eq:edge_cost} and the corresponding sum for edges in $Y$ is given by the third line of \eqref{eq:edge_cost}. As this type of edge mismatch cost involving an unassigned node is half the edge cost for a pair of assigned nodes, we refer to it as half-edge mismatch cost. 
    
\end{itemize}

In the graph GOSPA metric, $c$, $\epsilon$ and $p$ are hyperparameters. We can change $c$ to control the penalty for node unassignments, and the sensitivity to node attribute error. We can also adjust the value of $\epsilon$ to change the sensitivity to the changes in connectivity between nodes. Hyperparameter $p$ controls how much outliers are penalised. These parameters should be chosen according to the application.

As we will see in Section \ref{sec:lp_graph_metric}, the introduction of the half-edge mismatch cost allows us to express the metric using binary assignment matrices. This representation is computationally convenient as it gives rise to the linear programming relaxation of the metric, computable in polynomial time. 

The proof that $d^{(c,\epsilon)}(\cdot,\cdot)$ is a metric is provided as a particular case of the proof of the linear programming relaxation metric in Proposition \ref{prop:graphBinlp}, in Section \ref{sec:lp_graph_metric}.

\subsection{Examples}\label{subsec:examples}

We use Figure \ref{fig:eg1} to illustrate how the graph GOSPA metric works. This figure shows a ground truth graph $X$ and four different graph estimates $Y$. For simplicity, we consider the condition $p=1$, $\epsilon \ll c$ and $\Delta \ll c$, with $\Delta$ shown in Figure \ref{fig:eg1}.

We first consider Figure \ref{fig:eg1a}, where we have a graph $X$ with nodes ${x_1,x_2,x_3}$ (left to right) and graph $Y$ with nodes ${y_1,y_2,y_3}$. The optimal assignments for the nodes in $X$ are $(\pi_1,\pi_2,\pi_3)=(1,2,3)$, which means that each node in $X$ has been assigned to a node in $Y$ based on the distance between them.
In Figure \ref{fig:eg1a}, every edge between nodes is correctly detected, therefore, there is only localisation error in Figure \ref{fig:eg1a} and the metric value is, $d^{(c,\epsilon)}(X,Y)=3\Delta$, with each pair of assigned nodes contributing with $\Delta$.

In Figure \ref{fig:eg1b}, the assignment vector for $X$ is also $(1,2,3)$. In this case, apart from the three localisation errors, there is also a missed edge between nodes $y_1$ and $y_3$ within $Y$, such that the edge mismatch error $\epsilon$. The metric value for Figure \ref{fig:eg1b} is $d^{(c,\epsilon)}(X,Y)=3\Delta+\epsilon$.

In Figure \ref{fig:eg1c}, we compare two graphs with different sizes. The optimal assignments for nodes in $X$ are $(1,2,0)$. That is, the last node $x_3$ is left unassigned, which contributes to the error of a missed node, $c/2$. In this case, the edge between the two assigned nodes in $Y$ is properly detected. However, the third node with edges connected the other two nodes in $X$ does not exist in $Y$. Therefore, we penalise each missed edge connected to the missed node with half edge mismatch penalty $\epsilon /2$. Thus, the total edge mismatch cost in Figure \ref{fig:eg1c} is $\epsilon/2+\epsilon/2=\epsilon$. As there are two pairs of assigned nodes, each with a localisation error of $\Delta$, the metric value is $d^{(c,\epsilon)}(X,Y)=\Delta+\Delta+c/2+\epsilon=2\Delta+c/2+\epsilon$.

In Figure \ref{fig:eg1d}, we have one unassigned node in graph $X$, and another node in $Y$ is also unassigned, as the distance $\delta$ between $x_3$ and $y_3$ is much larger than $c$ and $\epsilon \ll c$. Therefore, $x_3$ a missed node and $y_3$ is a false node. The assignments are $(1,2,0)$. Each edge between the unassigned nodes and the assigned nodes contributes with a half-edge mismatch cost. As there are 3 of these edges, the total edge mismatch cost in Figure \ref{fig:eg1d} is $ \epsilon/2+ \epsilon/2+\epsilon/2=3\epsilon/2$. Then, as there are two pairs of assigned nodes, each with a localisation error of $\Delta$, the metric value is $d^{(c,\epsilon)}(X,Y)=\Delta+\Delta+c/2+c/2+3\epsilon/2=2\Delta+c+3\epsilon/2$.
According to the metric, the best estimate of $X$ is in Figure \ref{fig:eg1a}, which only has localisation errors. 
\tikzset{x=1.6 cm,y=1.6 cm, every text node part/.style={align=center}, every node/.style={font=\scriptsize, inner sep=1pt,outer sep=0pt, minimum size=2pt,  draw}}
\begin{figure}[t!]
\begin{center}
\begin{minipage}[t]{0.24\textwidth}
\centering
\begin{tikzpicture}
    \def \len {5}; 	\def \lenn {5};\def \del{0.35}; \def \ylim {-0.5};
    \node [rectangle, label =left:$X$](x1) at (1, 0){};
    \node [rectangle](x2) at (1.5,0){};
    \node [rectangle](x3) at (2,1){};
    \draw (x1) node[draw=none] at (1.05, -0.1) {$x_1$} --(x2) node[draw=none] at (1.4, -0.1) {$x_2$};
    \draw (x2)--(x3) node[draw=none] at (1.85, 1) {$x_3$};
    \draw (x1)--(x3);
	\node [circle, label = left:$Y$](y11) at (1, -\del){} node[draw=none] at (1.05, -\del-0.1){$y_1$};
	\node [circle](y12) at (1.5,-\del){} node[draw=none] at (1.5, -\del-0.1){$y_2$};
	\node [circle](y13) at (2,1-\del){} node[draw=none] at (2.1, 1-\del-0.1){$y_3$};
	\foreach \x in {2,..., 3} {
		\pgfmathtruncatemacro{\cur}{\x}
		\pgfmathtruncatemacro{\next}{\x - 1}
	\draw [densely dotted](y1\cur)--(y1\next);}
	\draw [densely dotted](y11)--(y13);
	\draw [<->]([xshift=5pt]x2.north west) -- ([xshift=5pt]y12.south west) node[draw=none,fill=none,midway,right] {$\Delta$};
	\draw [<->]([xshift=5pt]x3.north west) -- ([xshift=5pt]y13.south west) node[draw=none,fill=none,midway,right] {$\Delta$};
 \end{tikzpicture}
  \subcaption{}
\label{fig:eg1a}
 \end{minipage}
 \begin{minipage}[t]{0.24\textwidth}
\centering
\begin{tikzpicture}
    \def \len {5}; 	\def \lenn {5};\def \del{0.35}; \def \ylim {-0.5};
    \node [rectangle, label =left:$X$](x1) at (1, 0){};
    \node [rectangle](x2) at (1.5,0){};
    \node [rectangle](x3) at (2,1){};
    \foreach \x in {2,..., 3} {
		\pgfmathtruncatemacro{\cur}{\x}
		\pgfmathtruncatemacro{\next}{\x - 1}
	\draw (x\cur)--(x\next);}
    \draw (x1)--(x3);
	\node [circle, label = left:$Y$](y11) at (1, -\del){};
	\node [circle](y12) at (1.5,-\del){};
	\node [circle](y13) at (2,1-\del){};
	\foreach \x in {2,..., 3} {
		\pgfmathtruncatemacro{\cur}{\x}
		\pgfmathtruncatemacro{\next}{\x - 1}
	\draw [densely dotted](y1\cur)--(y1\next);}
	\draw [<->]([xshift=5pt]x2.north west) -- ([xshift=5pt]y12.south west) node[draw=none,fill=none,midway,right] {$\Delta$};
	\draw [<->]([xshift=5pt]x3.north west) -- ([xshift=5pt]y13.south west) node[draw=none,fill=none,midway,right] {$\Delta$};
\end{tikzpicture}
\subcaption{}
\label{fig:eg1b}
\end{minipage}
\begin{minipage}[t]{0.24\textwidth}
\centering
\begin{tikzpicture}
    \def \len {5}; 	\def \lenn {5};\def \del{0.35}; \def \ylim {-0.5};
    \node [rectangle, label =left:$X$](x1) at (1, 0){};
    \node [rectangle](x2) at (1.5,0){};
    \node [rectangle](x3) at (2,1){};
    \foreach \x in {2,..., 3} {
		\pgfmathtruncatemacro{\cur}{\x}
		\pgfmathtruncatemacro{\next}{\x - 1}
	\draw (x\cur)--(x\next);}
    \draw (x1)--(x3);
	\node [circle, label = left:$Y$](y11) at (1, -\del){};
	\node [circle](y12) at (1.5,-\del){};
%
	\foreach \x in {2} {
		\pgfmathtruncatemacro{\cur}{\x}
		\pgfmathtruncatemacro{\next}{\x - 1}
	\draw [densely dotted](y1\cur)--(y1\next);}
	\draw [<->]([xshift=5pt]x2.north west) -- ([xshift=5pt]y12.south west) node[draw=none,fill=none,midway,right] {$\Delta$};
%
%
 \end{tikzpicture}
  \subcaption{}
\label{fig:eg1c}
 \end{minipage}
\begin{minipage}[t]{0.24\textwidth}
\centering
\begin{tikzpicture}
    \def \len {5}; 	\def \lenn {5};\def \del{0.35}; \def \ylim {-0.5};
    \node [rectangle, label =left:$X$](x1) at (1, 0){};
    \node [rectangle](x2) at (1.5,0){};
    \node [rectangle](x3) at (2,1){};
    \foreach \x in {2,..., 3} {
		\pgfmathtruncatemacro{\cur}{\x}
		\pgfmathtruncatemacro{\next}{\x - 1}
	\draw (x\cur)--(x\next);}
    \draw (x1)--(x3);
	\node [circle, label = left:$Y$](y11) at (1, -\del){};
	\node [circle](y12) at (1.5,-\del){};
	\node [circle](y13) at (2,1-2*\del){};
	\foreach \x in {2,..., 3} {
		\pgfmathtruncatemacro{\cur}{\x}
		\pgfmathtruncatemacro{\next}{\x - 1}
	\draw [densely dotted](y1\cur)--(y1\next);}
	\draw [<->]([xshift=5pt]x2.north west) -- ([xshift=5pt]y12.south west) node[draw=none,fill=none,midway,right] {$\Delta$};
	\draw [<->]([xshift=5pt]x3.north west) -- ([xshift=5pt]y13.south west) node[draw=none,fill=none,midway,right] {$\delta$};
 \end{tikzpicture}
  \subcaption{}
\label{fig:eg1d}
 \end{minipage}
\caption{Example to illustrate the node and edge mismatch costs for the same ground truth graph $X$, and different estimated graphs $Y$ ($\Delta \ll c$): (a) three properly assigned nodes and no edge mismatch; (b) three properly assigned nodes and one missed edge; (c) two properly assigned nodes, one missed node and two  half-edge mismatch penalties; (d) Two properly assigned nodes and two unassigned nodes ($\delta \gg  c$), three half-edge mismatch penalties.  } 
\label{fig:eg1}
\vspace{-0.5cm}
 \end{center}
 \end{figure}
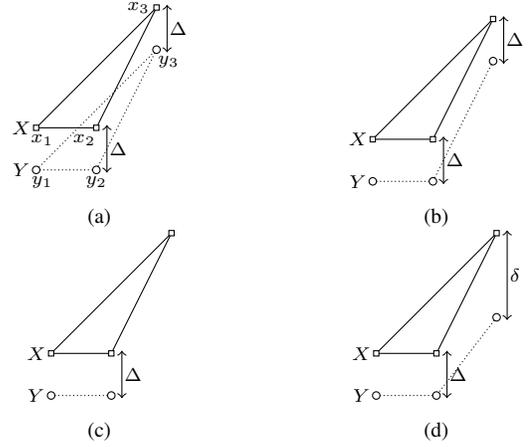

\section{LP Graph GOSPA metric}\label{sec:lp_graph_metric}

In this section, we show that the metric in (\ref{eq:graph_metric}) can be reformulated as an integer linear programming problem in Section \ref{lp formulation}. In Section \ref{lp relaxation}, we also show that the resulting distances are metrics for graphs. In Section \ref{sec:metric decomposition}, we explain the metric decomposition.

\subsection{Integer linear programming formulation}\label{lp formulation}

We proceed to write the graph GOSPA metric in terms of binary matrices instead of the assignment vectors in Section \ref{lp formulation}.
We use $\mathcal{W}_{X,Y}$ to denote the set of all binary matrices
of dimension $(n_{X}+1)\times(n_{Y}+1)$, representing assignments
between $V_{X}$ and $V_{Y}$. A matrix $W\in\mathcal{W}_{X,Y}$ satisfies
the following properties \cite{Angel20_d}:
\begin{flalign}
\sum_{i=1}^{n_{X}+1}W(i,j) &=1,\ j=1,\ldots,n_{Y}\label{eq:binary_constraint1}\\
\sum_{j=1}^{n_{Y}+1}W(i,j) &=1,\ i=1,\ldots,n_{X}\label{eq:binary_constraint2}\\
W(n_{X}+1,&n_{Y}+1) =0,\label{eq:binary_constraint3}\\
&W(i,j)  \in\{0,1\}, \, \forall\ i,j\label{eq:binary_constraint4}
\end{flalign}
where $W(i,j)$ is the element in row $i$ and column $j$ of matrix $W$. The element $W(i,j)=1$ if $x_i$ is assigned to $y_j$. If $x_i$ remains unassigned, $W(i,n_Y+1)=1$, and if $y_j$ remains unassigned then $W(n_X+1,j)=1$.

There is a bijection between the sets of binary matrices $\W_{X,Y}$ meeting the conditions in {\eqref{eq:binary_constraint1}--\eqref{eq:binary_constraint4}} and the sets of assignment vectors $\Pi_{V_X,V_Y}$, such that for $\pi \in \Pi_{X,Y}$, $W \in \W_{X,Y}$, $i=1,\hdots,n_X$ and $j=1,\hdots,n_Y$:
\begin{flalign}
&\pi_i=j \neq 0 &\Longleftrightarrow &W(i,j)=1\label{eq:pijw}\\
&\pi_i=0 &\Longleftrightarrow &W(i,n_Y+1)=1\label{eq:pi0winy+1}\\
&\nexists i \in \{1,\hdots,n_X\}, \pi_i=j \neq 0 &\Longleftrightarrow &W(n_X+1,j)=1.\label{eq:pijnx+1}
\end{flalign}

We also introduce the following matrix that contains all possible costs (to the $p$-th power) between pairs of nodes (including unassigned nodes).

Matrix $D_{X,Y}$ is an $(n_{X}+1)\times(n_{Y}+1)$ matrix whose $(i,j)$
element is:
\begin{flalign}\label{eq:Dxy}
D_{X,Y}(i,j) =\begin{cases}
d\left(x_{i},y_{j}\right)^{p}\ & i\leq n_{X},j\leq n_{Y},\\
\frac{c^{p}}{2} & i=n_{X}+1,j\leq n_{Y},\\
\frac{c^{p}}{2} & i\leq n_{X},j=n_{Y}+1,\\
0 & i=n_{X}+1,j=n_{Y}+1.
\end{cases}
\end{flalign}
The entries for $i \leq n_X$ and $j \leq n_Y$ represent localisation costs (to the $p$-th power) for assigned nodes. The entries for $i=n_X+1, j\leq n_Y$ and $j=n_Y+1, i\leq n_X$ represent the costs (to the $p$-th power) for unassigned nodes (miss/false detected nodes).

We also denote the component-wise 1-norm of a matrix $A\in \mathbb{R}^{n_X \times n_Y}$ as \cite{zhang2017matrix}:
\begin{flalign}
||A|| = \sum_{i=1}^{n_X}\sum_{j=1}^{n_Y} |A(i,j)|.
\end{flalign}
Then we can introduce Lemma \ref{lem:graph_gospa}.
\begin{lem}\label{lem:graph_gospa}
For $1<p<\infty$, a scalar $c > 0$, edge mismatch penalty $\epsilon>0$ and base metric $d(\cdot,\cdot)$, the graph GOSPA metric $d^{(c,\epsilon)}(\cdot,\cdot)$ in \eqref{eq:graph_metric} between two graphs $X$ and $Y$ can be
written as
\begin{flalign}
d^{\left(c,\epsilon\right)}(X,Y)
&=\min_{\substack{W\in\mathcal{W}_{X,Y}}}\left(\mathrm{tr}\big[D_{X,Y}^{\top}W\big] + e_{X,Y}(W)^p\right)^{1/p}\label{eq:simplified_graph_metric}
\end{flalign}
where 
\begin{flalign} \label{eq:edge_cost_mat}
    e_{X,Y}(W)^p= \frac{\epsilon^{p}}{2} || A_{X}W_{1:n_{X},1:n_{Y}} - W_{1:n_{X},1:n_{Y}}A_{Y}||
\end{flalign}
and $W_{1:n_{X},1:n_{Y}}$ is the matrix formed by the first $n_{X}$ rows and the first $n_{Y}$ columns of matrix $W$ (e.g., removing the last row and column of $W$).
\end{lem}

It should be noted that
\begin{flalign}
 \mathrm{tr}\big[D_{X,Y}^{\top}W\big] & =\sum_{i=1}^{n_{X}+1}\sum_{j=1}^{n_{Y}+1}D_{X,Y}(i,j)W\left(i,j\right)
\end{flalign}
and 
\begin{flalign}\label{eq:edge_cost_no_epsilon}
&|| A_{X}W_{1:n_{X},1:n_{Y}} - W_{1:n_{X},1:n_{Y}}A_{Y}|| \nonumber \\
&= \sum_{i=1}^{n_{X}}\sum_{j=1}^{n_{Y}}\left|\sum_{k=1}^{n_{X}}A_{X}(i,k)W(k,j) -\sum_{k=1}^{n_{Y}}W(i,k)A_{Y}(k,j)\right|.
\end{flalign}
The proof of Lemma \ref{lem:graph_gospa} is given in Appendix \ref{sec:proof of lemma 1}. Thus, we can write the optimisation problem based on the optimisation over assignment vectors $\pi \in \Pi_{X,Y}$ in \eqref{eq:graph_metric} in terms of an optimisation problem over assignment matrices $W\in \mathcal{W}_{X,Y}$ in \eqref{eq:simplified_graph_metric}.

If we set $\epsilon=0$ in \eqref{eq:simplified_graph_metric}, we obtain $d^{\left(c,0\right)}(\cdot,\cdot)$, which corresponds to the GOSPA metric \eqref{eq:gospa metric} applied to the set of nodes and can be efficiently computed by solving a 2-D assignment problem \cite{Rahmathullah17,Crouse16}. This actually provides a lower bound of $d^{\left(c,\epsilon\right)}(\cdot,\cdot)$. 
However, $d^{\left(c,0\right)}(\cdot,\cdot)$ is not a metric for graphs as edge information is discarded and therefore the identity property of metrics is not met.

\subsection{LP relaxation}\label{lp relaxation}
In this section, we relax the constraints of matrix $W$ in Lemma \ref{lem:graph_gospa} for polynomial computation time with linear programming, and we also show that the result is a metric.

Let $\hW_{X,Y}$ be the set of matrices of dimension $(n_X+1) \times (n_Y+1)$ with relaxation of the constraints in \eqref{eq:binary_constraint4} such that $W \in \hW_{X,Y}$ satisfies \eqref{eq:binary_constraint1}, \eqref{eq:binary_constraint2}, \eqref{eq:binary_constraint3} and
\begin{flalign} \label{eq:binary_constraint5}
    W(i,j) \geq 0,\, \forall i, j.
\end{flalign}

With the relaxation of the constraints in \eqref{eq:binary_constraint4}, we can write the metric function $\dS(X,Y)$ in \eqref{eq:simplified_graph_metric} as an alternative metric function $\dSh(X,Y)$, where the optimisation is over $W$ in $\hW_{X,Y}$ rather than $\W_{X,Y}$. Therefore it follows immediately that $\dSh(X,Y) \leq  \dS(X,Y)$.
\begin{prop}\label{prop:graphBinlp}
For $1 \leq p \leq \infty$, a scalar $c>0$, edge mismatch penalty $\epsilon>0$ and base metric $d(\cdot,\cdot)$, the LP relaxation of metric $d^{(c,\epsilon)(X,Y)}$ in \eqref{eq:simplified_graph_metric} is also a metric given by
\begin{flalign}\label{eq:LP graph metric}
\Bar{d}^{\left(c,\epsilon\right)}\left(X,Y\right) & =\min_{\substack{W\in\hW_{X,Y}}
}\left(\mathrm{tr}\big[D_{X,Y}^{\top}W\big]+e_{X,Y}(W)^p\right)^{1/p}
\end{flalign}
where $D_{X,Y}$ is given by \eqref{eq:Dxy} and $e_{X,Y}(W)^p$ is given by \eqref{eq:edge_cost_mat}.
\end{prop}

The proof of Proposition \ref{prop:graphBinlp} is given in Appendix \ref{proof:triangle undirected}. This proof also implies that the metric in \eqref{eq:simplified_graph_metric} is a metric.
In Appendix \ref{proof:LP proof}, we also prove that the metric can be computed using LP. This implies that there are optimisation algorithms that are ensured to converge and that the metric can be computed in polynomial time \cite{luenberger15}.

For Equation \eqref{eq:LP graph metric} in Proposition \ref{prop:graphBinlp}, set $\overline{\mathcal{W}}_{X, Y}$ is obtained by relaxing the binary constraints of set $\mathcal{W}_{X, Y}$ to $W(i,j) \geq 0, \forall i,j$. With the three other constraints in equation (4), (5), (6) in Section IV-A, the set $\overline{\mathcal{W}}_{X, Y}$ is closed and bounded. Since the metric function in equation (18) is continuous over $\overline{\mathcal{W}}_{X, Y}$, the minimum is achievable \cite[Thm. 4.6]{rudin1976}.

\subsection{Metric decomposition}\label{sec:metric decomposition}

In this section, we explain how the graph GOSPA metric, as shown in Lemma \ref{lem:graph_gospa}, and the LP graph GOSPA metric decompose into costs for properly assigned nodes, missed and false nodes, and edge mismatch costs. 

As explained after Definition \ref{defn:graph_gospa}, the node assignments are determined by the assignment vector $\pi$ defined in Section \ref{section:gospa_defn}. Missed nodes are those nodes in $V_X$ that are not assigned to a node in $V_Y$ according to $\pi$. False nodes are those nodes in $V_Y$ that are not assigned to a node in $V_X$ according to $\pi$. Then, $D_{X,Y}(i,j)$ represents the following costs:
\begin{itemize}
    \item A localisation error for assigned nodes if $i\leq n_X,$ $ j\leq n_Y$.
    \item A missed node if  $i \leq n_x,$ $ j=n_y +1$.
    \item A false node if $i = n_x +1,$ $ j \leq n_y $.
\end{itemize}

We denote the sets of indices $(i,j)$ that belong to each of the previous categories as $S_1$, $S_2$, $S_3$. Then, we have

\begin{flalign}
    d_p^{(c,\epsilon)}(X,Y) 
    &=\min_{\substack{W\in\mathcal{W}_{X,Y}}} \bigg( \mathrm{l}(X,Y,W)^p  
    + \mathrm{m}(X,Y,W)^p \nonumber \\
    &+ \mathrm{f}(X,Y,W)^p  
      + e_{X,Y}(W)^p \bigg)^{1/p} 
\end{flalign}
where 
\begin{flalign}
    \mathrm{l}(X,Y,W)^p &= \sum_{(i,j)\in S_1} D_{X,Y}(i,j) W(i,j) \nonumber \\
    \mathrm{m}(X,Y,W)^p &= \frac{c^p}{2} \sum_{(i,j)\in S_2} W(i,j) \nonumber \\
    \mathrm{f}(X,Y,W)^p &=  \frac{c^p}{2} \sum_{(i,j)\in S_3} W(i,j) \nonumber \\
\end{flalign}
represent the costs (to the $p$-th  power) for assigned, missed and false nodes given the assignment $W$. Therefore, we can decompose the metric in terms of these costs (and the edge mismatch cost) once we obtain the optimal assignment $W$. For the LP metric, we have the same decomposition but with soft assignments.

\section{Extensions to other types of graphs}\label{sec:extesion to weighted and directed graphs}
In the previous sections, we have presented the graph GOSPA metric for undirected, unweighted graphs, as the counting procedure to evaluate the graph edge mismatches is based on this type of graph. In Section \ref{sec:weighted graphs}, we explain that the metric in terms of binary matrices, see Lemma \ref{lem:graph_gospa} and Proposition \ref{prop:graphBinlp}, can be directly extended to weighted graphs by using the right forms of the adjacency matrices. In Section \ref{sec:directed graphs}, we show how the graph GOSPA metric can be extended to directed graphs, weighted and unweighted. In Section \ref{sec:no attributes}, we explain how the graph GOSPA metric can also be applied to graphs without node attributes, resulting in a pseudo-metric.

\subsection{Graph GOSPA metric for weighted undirected graphs}\label{sec:weighted graphs}
In a weighted undirected graph, the adjacency matrix is symmetric but is not constrained to have elements that are either 0 or 1. Each element now contains a non-zero scalar associated with its edge. For example, we can use the weight between edges to represent the type of bond between two atoms in a molecule. Note that, for the edge mismatch cost in \eqref{eq:edge_cost_mat}, a lack of edge would be equivalent to an edge with zero weight. Therefore, we assume that the edge weights are different from zero so that the identity property of metrics holds for weighted graphs.
For weighted graphs (with possible self-loops), we can establish the following proposition.
\begin{prop}
    The metrics $d^{(c,\epsilon)}(\cdot,\cdot)$ and $\Bar{d}^{\left(c,\epsilon\right)}(\cdot,\cdot)$ in \eqref{eq:graph_metric} and \eqref{eq:LP graph metric}, respectively, are also metrics for weighted undirected graphs (with possible self-loops).
\end{prop}

The proof is the same as the one in Appendix \ref{sec:proof of triangle inequality}, as it is also valid for weighted and undirected graphs. The main difference with undirected, unweighted graphs is that now the adjacency matrix elements are not constrained to be 0 or 1. 

The interpretation of the graph GOSPA metric for undirected, weighted graphs is the same as for undirected, unweighted graphs in terms of node attribute errors, missed and false nodes, but differs in the edge mismatch penalty, which now considers the differences in the edge weights.
The computational complexity of the LP implementation of the graph GOSPA metric for weighted graphs is similar to the case of unweighted graphs, as there are the same number of variables and constraints.

We use Figure \ref{fig:eg2} to illustrate how the graph GOSPA metric works on weighted graphs. Figure \ref{fig:eg2} considers weighted versions of the graphs in Figure \ref{fig:eg1}. As in the examples in Figure \ref{fig:eg1}, we consider $p=1$, $\epsilon \ll c$, and $\Delta \ll c$, and also that $\Delta$ is sufficiently small such that the assignment matrices with respect to Figure \ref{fig:eg1} do not change. In this case, the localisation errors remain unchanged and we only proceed to calculate the edge mismatch costs. 

In Figure \ref{fig:eg2a}, there is no actual edge mismatch, but there is a weight difference between graph $X$ and graph $Y$. Therefore, the edge mismatch cost is $e_{X,Y}=\epsilon \times (0.7-0.4)=0.3\epsilon$, and the metric value is $d^{(c,\epsilon)}(X,Y)=3\Delta+0.3\epsilon$. 

In Figure \ref{fig:eg2b}, we also have a missed edge between nodes $y_1$ and $y_2$ in $Y$, and the weight of the corresponding edge in $X$ is 0.7, so the edge mismatch error for this case is $0.7\epsilon$.
The metric value for Figure \ref{fig:eg2b} is $d^{(c,\epsilon)}(X,Y)=3\Delta+0.7\epsilon$.

In Figure \ref{fig:eg2c}, as we have two half-edge mismatch penalties, and the weights for the missed edges are 0.7 and 0.5. The edge mismatch error is then $e_{X,Y}=\epsilon/2 \times (0.7+0.5)=0.6\epsilon$. The metric value is $d^{(c,\epsilon)}(X,Y)=2\Delta+c/2+0.6\epsilon$.

In Figure \ref{fig:eg2d}, similarly, there are three half-edge mismatch penalties, and the weight for each edge is 0.7, 0.5, 0.6 respectively. The edge mismatch error is $e_{X,Y}=\epsilon/2 \times (0.7+0.5+0.6)=0.9\epsilon$. The metric value is $d^{(c,\epsilon)}(X,Y)=2\Delta+c+0.9\epsilon$.
\tikzset{x=1.6 cm,y=1.6 cm, every text node part/.style={align=center}, every node/.style={font=\scriptsize, inner sep=1pt,outer sep=0pt, minimum size=2pt}}
\begin{figure}[t!]
\begin{center}
\begin{minipage}[t]{0.24\textwidth}
\centering
\begin{tikzpicture}
    \def \len {5}; 	\def \lenn {5};\def \del{0.35}; \def \ylim {-0.5};
    \node [draw,rectangle, label =left:$X$](x1) at (1, 0){}
    node[draw=none] at (1.05, -0.1) {$x_1$};
    \node [draw,rectangle](x2) at (1.5,0){}
    node[draw=none] at (1.48, -0.1) {$x_2$};
    \node [draw,rectangle](x3) at (2,1){}
    node[draw=none] at (1.85, 1) {$x_3$};
    
    \draw (x1)--(x2) node[midway,below,black]{$0.3$};
    \draw (x2)--(x3) node[pos=0.4,right,black]{$0.5$};
    \draw (x1)--(x3) node[pos=0.5,left,black]{$0.7$};

    \node [draw,circle, label = left:$Y$](y11) at (1, -\del){}
        node[draw=none] at (1, -\del-0.1) {$y_1$};
    \node [draw,circle](y12) at (1.5,-\del){}
        node[draw=none] at (1.5, -\del-0.1) {$y_2$};
    \node [draw,circle](y13) at (2,1-\del){}
        node[draw=none] at (2.1, 1-\del-0.1) {$y_3$};
%
        \draw [densely dotted](y12)--(y13) node[midway,below,text=red]{$0.5$};
        \draw [densely dotted](y11)--(y12) node[midway,below,text=red]{$0.3$};
	\draw [densely dotted](y11)--(y13) node[pos=0.4,above,text=red]{$0.4$};
	\draw [<->]([xshift=5pt]x2.north west) -- ([xshift=5pt]y12.south west) node[draw=none,fill=none,midway,right] {$\Delta$};
	\draw [<->]([xshift=5pt]x3.north west) -- ([xshift=5pt]y13.south west) node[draw=none,fill=none,midway,right] {$\Delta$};
 \end{tikzpicture}
  \subcaption{}
\label{fig:eg2a}
 \end{minipage}
 \begin{minipage}[t]{0.24\textwidth}
\centering
\begin{tikzpicture}
    \def \len {5}; 	\def \lenn {5};\def \del{0.35}; \def \ylim {-0.5};
    \node [draw,rectangle, label =left:$X$](x1) at (1, 0){};
    \node [draw,rectangle](x2) at (1.5,0){};
    \node [draw,rectangle](x3) at (2,1){};
    \draw (x1)--(x2) node[midway,below,black]{$0.3$};
    \draw (x2)--(x3) node[pos=0.4,right,black]{$0.5$};
    \draw (x1)--(x3) node[pos=0.5,left,black]{$0.7$};
	\node [draw,circle, label = left:$Y$](y11) at (1, -\del){};
	\node [draw,circle](y12) at (1.5,-\del){};
	\node [draw,circle](y13) at (2,1-\del){};
	\draw [densely dotted](y12)--(y13) node[midway,below,text=red]{$0.5$};
        \draw [densely dotted](y11)--(y12) node[midway,below,text=red]{$0.3$};
	\draw [<->]([xshift=5pt]x2.north west) -- ([xshift=5pt]y12.south west) node[draw=none,fill=none,midway,right] {$\Delta$};
	\draw [<->]([xshift=5pt]x3.north west) -- ([xshift=5pt]y13.south west) node[draw=none,fill=none,midway,right] {$\Delta$};
\end{tikzpicture}
\subcaption{}
\label{fig:eg2b}
\end{minipage}
\begin{minipage}[t]{0.24\textwidth}
\centering
\begin{tikzpicture}
    \def \len {5}; 	\def \lenn {5};\def \del{0.35}; \def \ylim {-0.5};
    \node [draw,rectangle, label =left:$X$](x1) at (1, 0){};
    \node [draw,rectangle](x2) at (1.5,0){};
    \node [draw,rectangle](x3) at (2,1){};
    \draw (x1)--(x2) node[midway,below,black]{$0.3$};
    \draw (x2)--(x3) node[pos=0.4,right,black]{$0.5$};
    \draw (x1)--(x3) node[pos=0.5,left,black]{$0.7$};
	\node [draw,circle, label = left:$Y$](y11) at (1, -\del){};
	\node [draw,circle](y12) at (1.5,-\del){};
%
	\draw [densely dotted](y11)--(y12) node[midway,below,text=red]{$0.3$};
	\draw [<->]([xshift=5pt]x2.north west) -- ([xshift=5pt]y12.south west) node[draw=none,fill=none,midway,right] {$\Delta$};
%
%
 \end{tikzpicture}
  \subcaption{}
\label{fig:eg2c}
 \end{minipage}
\begin{minipage}[t]{0.24\textwidth}
\centering
\begin{tikzpicture}
    \def \len {5}; 	\def \lenn {5};\def \del{0.35}; \def \ylim {-0.5};
    \node [draw,rectangle, label =left:$X$](x1) at (1, 0){};
    \node [draw,rectangle](x2) at (1.5,0){};
    \node [draw,rectangle](x3) at (2,1){};
    \draw (x1)--(x2) node[midway,below,black]{$0.3$};
    \draw (x2)--(x3) node[pos=0.4,right,black]{$0.5$};
    \draw (x1)--(x3) node[pos=0.5,left,black]{$0.7$};
    
	\node [draw,circle, label = left:$Y$](y11) at (1, -\del){};
	\node [draw,circle](y12) at (1.5,-\del){};
	\node [draw,circle](y13) at (2,1-2*\del){};

        \draw [densely dotted](y12)--(y13) node[pos=0.7,below,text=red]{$0.6$};
        \draw [densely dotted](y11)--(y12) node[midway,below,text=red]{$0.3$};
	\draw [<->]([xshift=5pt]x2.north west) -- ([xshift=5pt]y12.south west) node[draw=none,fill=none,midway,right] {$\Delta$};
	\draw [<->]([xshift=5pt]x3.north west) -- ([xshift=5pt]y13.south west) node[draw=none,fill=none,midway,right] {$\delta$};
 \end{tikzpicture}
  \subcaption{}
\label{fig:eg2d}
 \end{minipage}
\caption{Example to illustrate the node and edge mismatch costs for weighted graphs, ground truth graph $X$, and different estimated graphs $Y$: (a) three properly assigned nodes, 0.3 edge-mismatch penalty; (b) three properly assigned nodes ($\Delta \ll c$) and one missed edge with 0.7 edge-mismatch penalty; (c) two properly assigned nodes, one missed node and 0.6 edge mismatch penalties; (d) Two properly assigned nodes ($\Delta \ll c$) and two unassigned nodes ($\delta \gg  c$), 0.9 edge mismatch penalties.  } 
\label{fig:eg2}
\vspace{-0.5cm}
 \end{center}
 \end{figure}
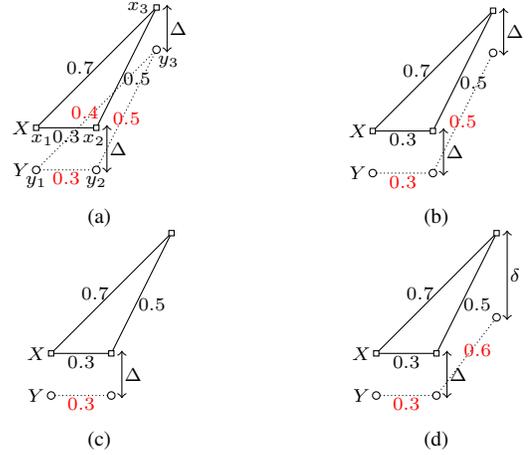

\subsection{Graph GOSPA metric for directed graphs}\label{sec:directed graphs}
In a directed unweighted graph, the adjacency matrix has elements that are either 0 or 1, but it is not symmetric. A possible application of directed graphs is modelling the one-way connectivities between users in social networks. In directed weighted graphs, the adjacency matrix has elements that can be any real number and it is not symmetric.
Due to the non-symmetric adjacency matrices in directed graphs, the metric for directed graphs must be adapted, as \eqref{eq:simplified_graph_metric} only meets the symmetry property for symmetric $A_X$ and $A_Y$. The extension of this metric for directed graphs (both weighted and unweighted) is given in the following lemma.

\begin{lem}\label{lem:extension directed graphs}

For $1 \leq p \leq \infty$, a scalar $c>0$, edge mismatch penalty $\epsilon>0$ and base metric $d(\cdot,\cdot)$, the graph GOSPA metric for directed graphs $X$ and $Y$ is 
\begin{flalign}\label{eq:graph gospa directed}
d^{\left(c,\epsilon\right)}(X,Y)
=\min_{\substack{W\in\mathcal{W}_{X,Y}}}\left(\mathrm{tr}\big[D_{X,Y}^{\top}W\big] + e_{X,Y}(W)^p\right)^{1/p} 
\end{flalign}
where $D_{X,Y}$ is given by \eqref{eq:Dxy} and the edge cost is
\begin{flalign}\label{eq:edge cost directed}
    e_{X,Y}(W)=&\frac{\epsilon^p}{4} \bigg(||A_X W_{1:n_X,1:n_Y}- W_{1:n_X,1:n_Y}A_Y||  \nonumber\\
     &+||A_Y W^T_{1:n_X,1:n_Y}- W^T_{1:n_X,1:n_Y}A_X|| \bigg) 
\end{flalign}

\end{lem}

The LP lower bound of \eqref{eq:graph gospa directed}, $\bar{d}^{(c,\epsilon)}(\cdot,\cdot)$ is also a metric and is obtained by minimising over $W \in \bar{W}_{X,Y}$.

Both $d^{(c,\epsilon)}(\cdot,\cdot)$ and $\bar{d}^{(c,\epsilon)}(\cdot,\cdot)$ for directed graphs are symmetric and meet the identity property. The proof of the triangle inequality is given in Appendix \ref{sec:triangle directed}. In Appendix \ref{proof:LP proof} we prove that $\bar{d}^{(c,\epsilon)}(\cdot,\cdot)$ for directed graphs is also computable in polynomial time using LP. Note that if $A_X$ and $A_Y$ are symmetric (as in undirected graphs), the edge mismatch cost \eqref{eq:edge cost directed} can be written as in \eqref{eq:edge_cost_mat}. Therefore, \eqref{eq:edge cost directed} is a general edge mismatch cost that can be used to cover directed and undirected graphs, and for undirected graphs, \eqref{eq:graph gospa directed} coincides with \eqref{eq:simplified_graph_metric}. The computational complexity of the LP implementation of the graph GOSPA metric for directed graphs is higher than the implementation for undirected graphs as there is a higher number of constraints, as explained in Appendix \ref{proof:LP proof} \cite{KHACHIYAN198053}.

The decomposition of the graph GOSPA metric for directed graphs into
node attribute errors, missed and false nodes are the same as for undirected
graphs. The edge mismatch cost is obtained as follows:
\begin{itemize}
\item For every two pairs of properly assigned nodes (two nodes in $X$
and two nodes in $Y$), if a (directed) edge exists in one graph
but it does not exist in the other graph, the associated edge mismatch
penalty is $\epsilon^p/2$.
\item Each (directed) edge in either graph connecting an assigned node and
an unassigned node contributes to an edge mismatch penalty $\epsilon^p/4$. 
\end{itemize}
We proceed to use the examples in Figure \ref{fig:eg3} to explain how the graph GOSPA metric works on directed graphs. We also use the same hyper-parameters $p=1$ and $\Delta \ll c$.

In Figure \ref{fig:eg3a}, the edge between nodes $x_1$ and $x_3$ in $X$ has the opposite direction to the edge between nodes $y_1$ and $y_3$ in $Y$, which yields one edge mismatch penalty, $\epsilon$. The metric value in this case is $\hat{d}^{(c,\epsilon)}(X,Y)=3\Delta+\epsilon$.

In Figure \ref{fig:eg3b}, there is a missed edge between nodes $y_1$ and $y_3$ in $Y$, and this contributes to a half-edge mismatch penalty, $\epsilon/2$. The total metric value is $\hat{d}^{(c,\epsilon)}(X,Y)=3\Delta+\epsilon/2$.

In Figure \ref{fig:eg3c}, there is a missed node and two missed edges, and each gives a quarter-edge mismatch penalty, $e_{X,Y}=\epsilon/2$. The metric value for Figure \ref{fig:eg3c} is $\hat{d}^{(c,\epsilon)}(X,Y)=2\Delta+c/2+\epsilon/2$.

In Figure \ref{fig:eg3d}, there are two unassigned nodes, two missed edges and one extra edge, and in this case, the edge error is $\epsilon$. The total metric value is $\hat{d}^{(c,\epsilon)}(X,Y)=2\Delta+c+3\epsilon/4$.
\tikzset{x=1.6 cm,y=1.6 cm, every text node part/.style={align=center}, every node/.style={font=\scriptsize, inner sep=1pt,outer sep=0pt, minimum size=2pt,  draw}}
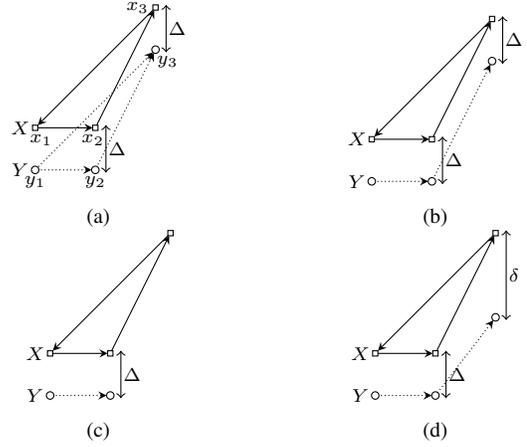
\begin{figure}[t!]
\begin{center}
\begin{minipage}[t]{0.24\textwidth}
\centering
\begin{tikzpicture}
    \def \len {5}; 	\def \lenn {5};\def \del{0.35}; \def \ylim {-0.5};
    \node [draw,rectangle, label =left:$X$](x1) at (1, 0){}
    node[draw=none] at (1.05, -0.1) {$x_1$};
    \node [draw,rectangle](x2) at (1.5,0){}
    node[draw=none] at (1.48, -0.1) {$x_2$};
    \node [draw,rectangle](x3) at (2,1){}
    node[draw=none] at (1.85, 1) {$x_3$};
    \foreach \x in {2,..., 3} {
		\pgfmathtruncatemacro{\cur}{\x}
		\pgfmathtruncatemacro{\next}{\x - 1}
	\draw [stealth-](x\cur)--(x\next);}
    \draw [stealth-](x1)--(x3);
    \node [draw,circle, label = left:$Y$](y11) at (1, -\del){}
        node[draw=none] at (1, -\del-0.1) {$y_1$};
    \node [draw,circle](y12) at (1.5,-\del){}
        node[draw=none] at (1.5, -\del-0.1) {$y_2$};
    \node [draw,circle](y13) at (2,1-\del){}
        node[draw=none] at (2.1, 1-\del-0.1) {$y_3$};
        
	\foreach \x in {2,..., 3} {
		\pgfmathtruncatemacro{\cur}{\x}
		\pgfmathtruncatemacro{\next}{\x - 1}
	\draw [stealth-][densely dotted](y1\cur)--(y1\next);}
	\draw [stealth-][densely dotted](y13)--(y11);
	\draw [<->]([xshift=5pt]x2.north west) -- ([xshift=5pt]y12.south west) node[draw=none,fill=none,midway,right] {$\Delta$};
	\draw [<->]([xshift=5pt]x3.north west) -- ([xshift=5pt]y13.south west) node[draw=none,fill=none,midway,right] {$\Delta$};
 \end{tikzpicture}
  \subcaption{}
\label{fig:eg3a}
 \end{minipage}
 \begin{minipage}[t]{0.24\textwidth}
\centering
\begin{tikzpicture}
    \def \len {5}; 	\def \lenn {5};\def \del{0.35}; \def \ylim {-0.5};
    \node [rectangle, label =left:$X$](x1) at (1, 0){};
    \node [rectangle](x2) at (1.5,0){};
    \node [rectangle](x3) at (2,1){};
    \foreach \x in {2,..., 3} {
		\pgfmathtruncatemacro{\cur}{\x}
		\pgfmathtruncatemacro{\next}{\x - 1}
	\draw [stealth-](x\cur)--(x\next);}
    \draw [stealth-](x1)--(x3);
	\node [circle, label = left:$Y$](y11) at (1, -\del){};
	\node [circle](y12) at (1.5,-\del){};
	\node [circle](y13) at (2,1-\del){};
	\foreach \x in {2,..., 3} {
		\pgfmathtruncatemacro{\cur}{\x}
		\pgfmathtruncatemacro{\next}{\x - 1}
	\draw [stealth-][densely dotted](y1\cur)--(y1\next);}
	\draw [<->]([xshift=5pt]x2.north west) -- ([xshift=5pt]y12.south west) node[draw=none,fill=none,midway,right] {$\Delta$};
	\draw [<->]([xshift=5pt]x3.north west) -- ([xshift=5pt]y13.south west) node[draw=none,fill=none,midway,right] {$\Delta$};
\end{tikzpicture}
\subcaption{}
\label{fig:eg3b}
\end{minipage}
\begin{minipage}[t]{0.24\textwidth}
\centering
\begin{tikzpicture}
    \def \len {5}; 	\def \lenn {5};\def \del{0.35}; \def \ylim {-0.5};
    \node [rectangle, label =left:$X$](x1) at (1, 0){};
    \node [rectangle](x2) at (1.5,0){};
    \node [rectangle](x3) at (2,1){};
    \foreach \x in {2,..., 3} {
		\pgfmathtruncatemacro{\cur}{\x}
		\pgfmathtruncatemacro{\next}{\x - 1}
	\draw [stealth-](x\cur)--(x\next);}
    \draw [stealth-](x1)--(x3);
	\node [circle, label = left:$Y$](y11) at (1, -\del){};
	\node [circle](y12) at (1.5,-\del){};
%
	\foreach \x in {2} {
		\pgfmathtruncatemacro{\cur}{\x}
		\pgfmathtruncatemacro{\next}{\x - 1}
	\draw [stealth-][densely dotted](y1\cur)--(y1\next);}
	\draw [<->]([xshift=5pt]x2.north west) -- ([xshift=5pt]y12.south west) node[draw=none,fill=none,midway,right] {$\Delta$};
%
%
 \end{tikzpicture}
  \subcaption{}
\label{fig:eg3c}
 \end{minipage}
\begin{minipage}[t]{0.24\textwidth}
\centering
\begin{tikzpicture}
    \def \len {5}; 	\def \lenn {5};\def \del{0.35}; \def \ylim {-0.5};
    \node [rectangle, label =left:$X$](x1) at (1, 0){};
    \node [rectangle](x2) at (1.5,0){};
    \node [rectangle](x3) at (2,1){};
    \foreach \x in {2,..., 3} {
		\pgfmathtruncatemacro{\cur}{\x}
		\pgfmathtruncatemacro{\next}{\x - 1}
	\draw [stealth-](x\cur)--(x\next);}
    \draw [stealth-](x1)--(x3);
	\node [circle, label = left:$Y$](y11) at (1, -\del){};
	\node [circle](y12) at (1.5,-\del){};
	\node [circle](y13) at (2,1-2*\del){};
	\foreach \x in {2,..., 3} {
		\pgfmathtruncatemacro{\cur}{\x}
		\pgfmathtruncatemacro{\next}{\x - 1}
	\draw [stealth-][densely dotted](y1\cur)--(y1\next);}
	\draw [<->]([xshift=5pt]x2.north west) -- ([xshift=5pt]y12.south west) node[draw=none,fill=none,midway,right] {$\Delta$};
	\draw [<->]([xshift=5pt]x3.north west) -- ([xshift=5pt]y13.south west) node[draw=none,fill=none,midway,right] {$\delta$};
 \end{tikzpicture}
  \subcaption{}
\label{fig:eg3d}
 \end{minipage}
\caption{Example to illustrate the node and edge mismatch costs for directed graphs, ground truth: $X$ and estimated: $Y$: (a) three properly assigned nodes and one edge mismatch penalty; (b) three properly assigned nodes ($\Delta \ll c$) and half edge mismatch penalty; (c) two properly assigned nodes, one missed node and half edge mismatch penalty; (d) Two properly assigned nodes ($\Delta \ll c$) and two unassigned nodes ($\delta \gg  c$), one edge mismatch penalty.  } 
\label{fig:eg3}
\vspace{-0.5cm}
 \end{center}
 \end{figure}

\subsection{Graphs without node attributes}\label{sec:no attributes}

There are applications in which graphs do not have node attributes with physical meaning, or these attributes are not available. In this case,
one can denote the set of nodes with integers, such that
$V_{X}=\left\{ 1,...,n_{X}\right\} $ and $V_{Y}=\left\{ 1,...,n_{Y}\right\} $.
Therefore, the localisation part of the graph GOSPA metric does not
have a physical meaning either. In this case, the base metric can
be set to zero $d\left(\cdot,\cdot\right)=0$ and the following lemma
holds.

\begin{lem}\label{lem:graph metric no attibutes}
If the base distance is $d\left(\cdot,\cdot\right)=0$, the function
in \eqref{eq:LP graph metric} becomes
\begin{flalign}\label{eq:LP graph metric no attributes}
\tilde{d}^{\left(c,\epsilon\right)}\left(X,Y\right) & =\min_{\substack{W\in\hW_{X,Y}}}
\left(\frac{c^{p}}{2} \left(\sum_{i=1}^{n_X}W(i,n_Y+1) \right. \right. \nonumber\\
 & \left.\left. +\sum_{j=1}^{n_Y}W(n_X+1,j) \right)+e_{X,Y}(W)^p \right)^{1/p}
\end{flalign}
where $W$ satisfies \eqref{eq:binary_constraint1}, \eqref{eq:binary_constraint2}, \eqref{eq:binary_constraint3} and \eqref{eq:binary_constraint5}, and $e_{X,Y}$ is given in \eqref{eq:edge_cost_mat} for undirected graphs and \eqref{eq:edge cost directed} for directed graphs.

The function $\tilde{d}^{(c,\epsilon)}\left(\cdot,\cdot\right)$ is
a pseudometric for graphs \cite{howes2012modern}, which implies:
\begin{enumerate}
    \item $\tilde{d}^{(c,\epsilon)}(X,X)=0$;
    \item (symmetry) $\tilde{d}^{(c,\epsilon)}(X,Y)=\tilde{d}^{(c,\epsilon)}(Y,X)$;
    \item (triangle inequality) $\tilde{d}^{(c,\epsilon)}(X,Y) \leq \tilde{d}^{(c,\epsilon)}(X,Z) + \tilde{d}^{(c,\epsilon)}(Z,Y)$.
\end{enumerate}
\end{lem}
The proof of Lemma \ref{lem:graph metric no attibutes} is given in Appendix \ref{sec:proof of prop3: metric without attributes} and also holds for the non-relaxed version of the graph GOSPA metric. Additionally, it is direct to check that $\tilde{d}^{(c,\epsilon)}\left(X,Y\right)=0$ if and only if $X$ and $Y$ are isomorphic. That is, graphs $X$ and $Y$ are isomorphic if and only if they are of the same size and there is a permutation matrix $P$ such that $A_{X}P =PA_{Y}$ \cite{Bento19}.

The pseudometric $\tilde{d}^{(c,\epsilon)}\left(\cdot,\cdot\right)$
then defines an equivalence relation in $\Omega$ such that $X\sim Y$
if $\tilde{d}^{(c,\epsilon)}\left(X,Y\right)=0$ \cite{howes2012modern}.
Then, the equivalence class of an element (graph) $X$, denoted by
$\left[X\right]$, is the set of elements in $\Omega$ equivalent
to $X$ such that $\left[X\right]  =\left\{ Y\in\Omega:Y\sim X\right\}$, which is the set of graphs $Y$ that are isomorphic to $X$. The set of equivalence classes of isomorphic graphs is referred to as quotient space $\Omega\left\backslash \sim\right.$ , such that $\left[X\right]\in\Omega\left\backslash \sim\right.$ \cite{howes2012modern}.

\begin{lem}
The distance $\tilde{\rho}^{(c,\epsilon)}\left(\cdot,\cdot\right)$
on the quotient space $\Omega\left\backslash \sim\right.$ (set of
equivalence classes of isomorphic graphs) defined as
\begin{align}
\tilde{\rho}^{(c,\epsilon)}\left(\left[X\right],\left[Y\right]\right) & =\tilde{d}^{(c,\epsilon)}\left(X,Y\right)
\end{align}
where $X$ and $Y$ are two elements of $\left[X\right]$ and $\left[Y\right]$,
respectively, is a metric on $\Omega\left\backslash \sim\right.$ \normalfont \cite{howes2012modern}.
\end{lem}

We now proceed to write the pseudo-metric $\tilde{d}^{(c,\epsilon)}\left(\cdot,\cdot\right)$
values for the graphs in Figure \ref{fig:eg1}, by removing their node attributes. 
In Figure \ref{fig:eg1a}, the two graphs are isomorphic so $\tilde{d}^{(c,\epsilon)}\left(X,Y\right)=0$. In Figure \ref{fig:eg1b}, there is only one edge mismatch cost $\epsilon$, therefore, $\tilde{d}^{(c,\epsilon)}\left(X,Y\right)=\epsilon$. In Figure \ref{fig:eg1c}, the graphs have different sizes, which contributes to a cardinality error $c/2$, and $\tilde{d}^{(c,\epsilon)}\left(X,Y\right)=c/2+\epsilon$. In Figure \ref{fig:eg1d}, as there are no node attributes in both graphs, the two graphs $X$ and $Y$ are the same as the graphs in Figure \ref{fig:eg1b}, therefore, they have the same metric value, $\tilde{d}^{(c,\epsilon)}\left(X,Y\right)=\epsilon$.

\begin{figure}[t]
    \centering
    \includegraphics[width=.9\linewidth]{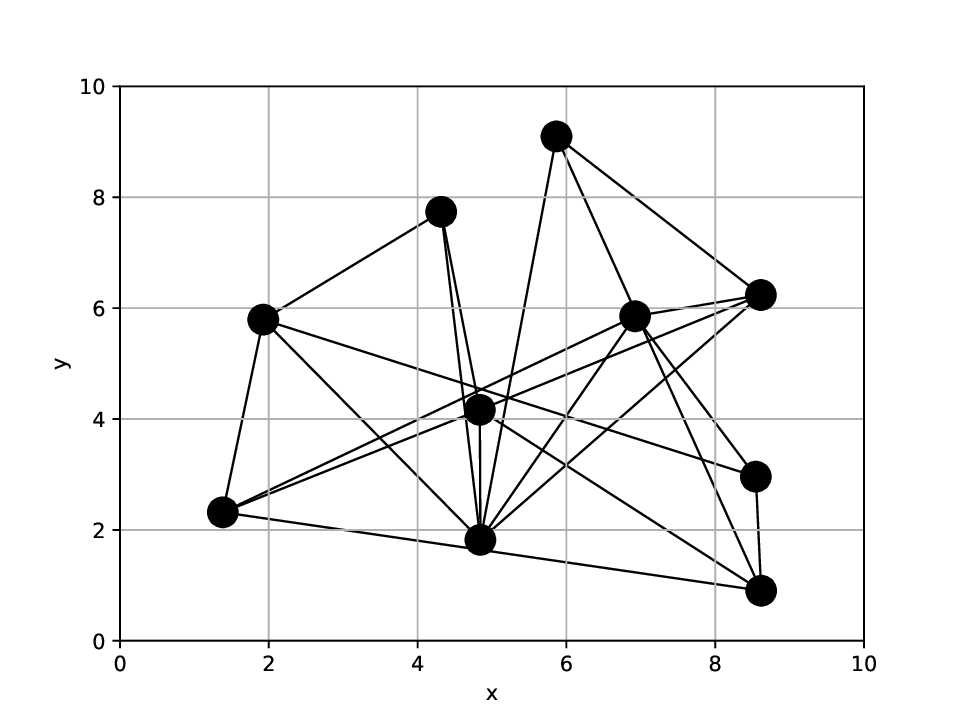}
    \caption{Randomly generated ground truth graph $X$.}
    \label{fig:synthetic graph}
\end{figure}

\begin{figure*}[t]
\begin{subfigure}{.5\textwidth}
  \centering
  \includegraphics[width=.8\linewidth]{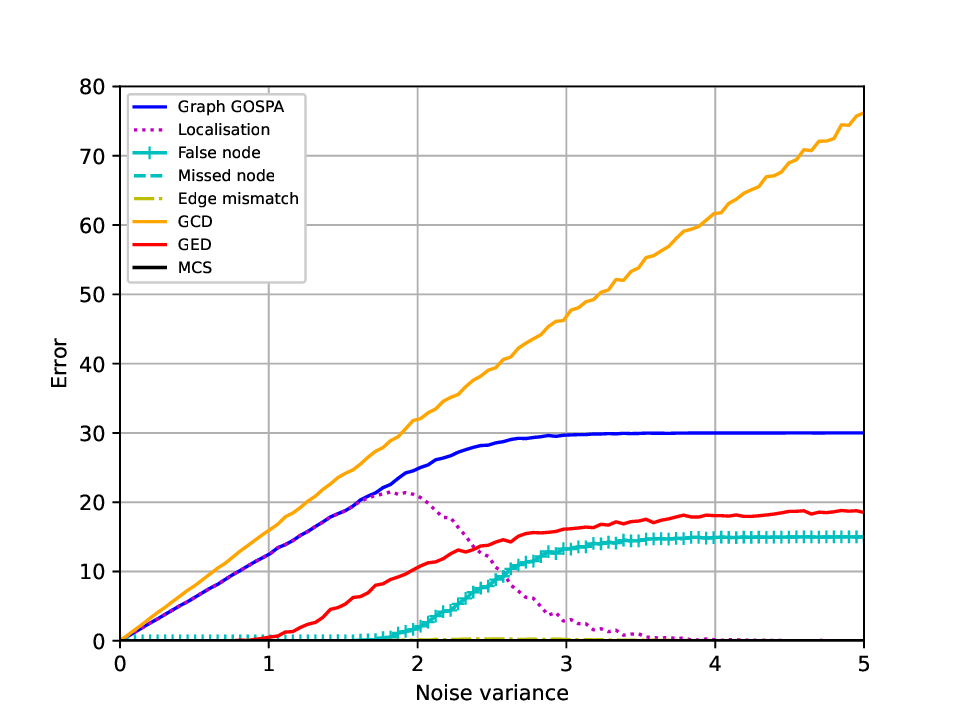}
  \caption{Random node attribute noise.}
  \label{fig:exp_a}
\end{subfigure}
\begin{subfigure}{.5\textwidth}
  \centering
  \includegraphics[width=.8\linewidth]{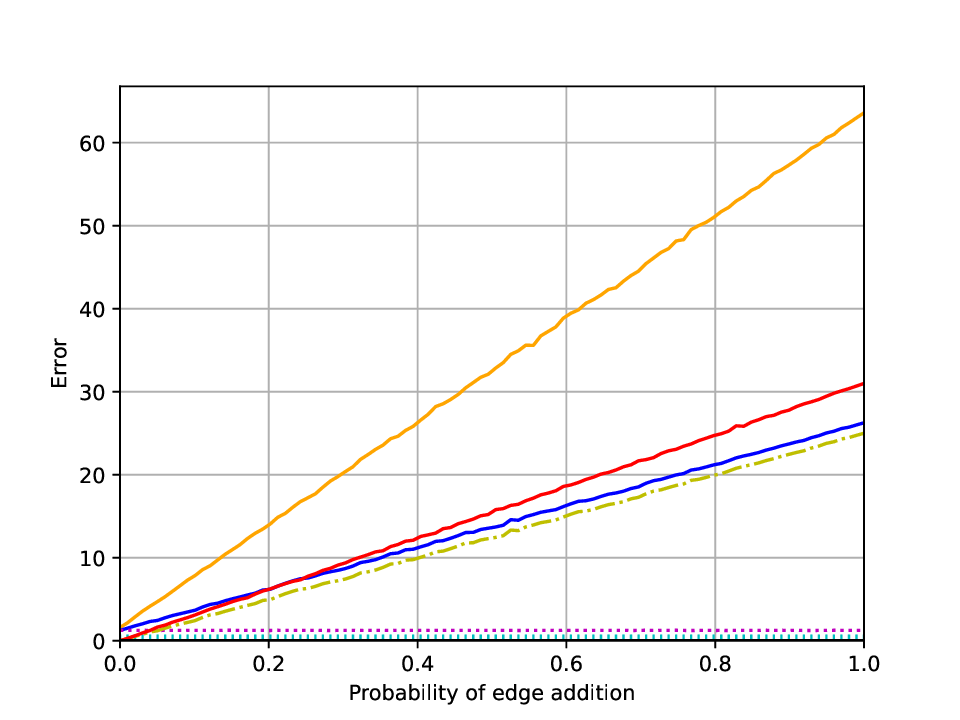}
  \caption{Random edge addition.}
  \label{fig:exp_b}
\end{subfigure}
\begin{subfigure}{.5\textwidth}
  \centering
  \includegraphics[width=.8\linewidth]{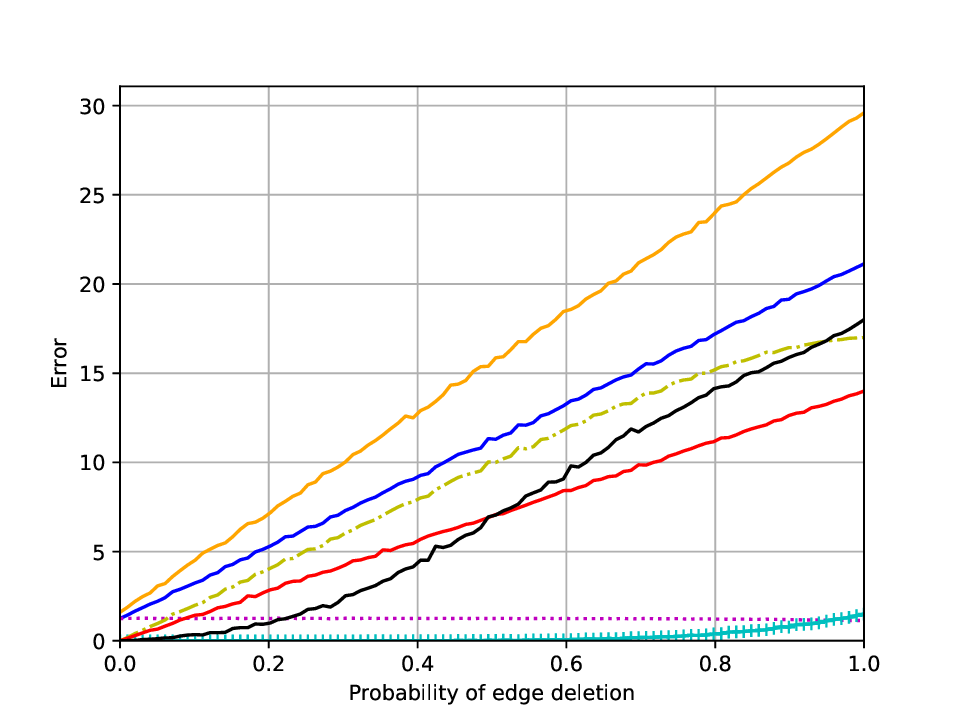}
  \caption{Random edge deletion.}
  \label{fig:exp_c}
\end{subfigure}
\begin{subfigure}{.5\textwidth}
  \centering
  \includegraphics[width=.8\linewidth]{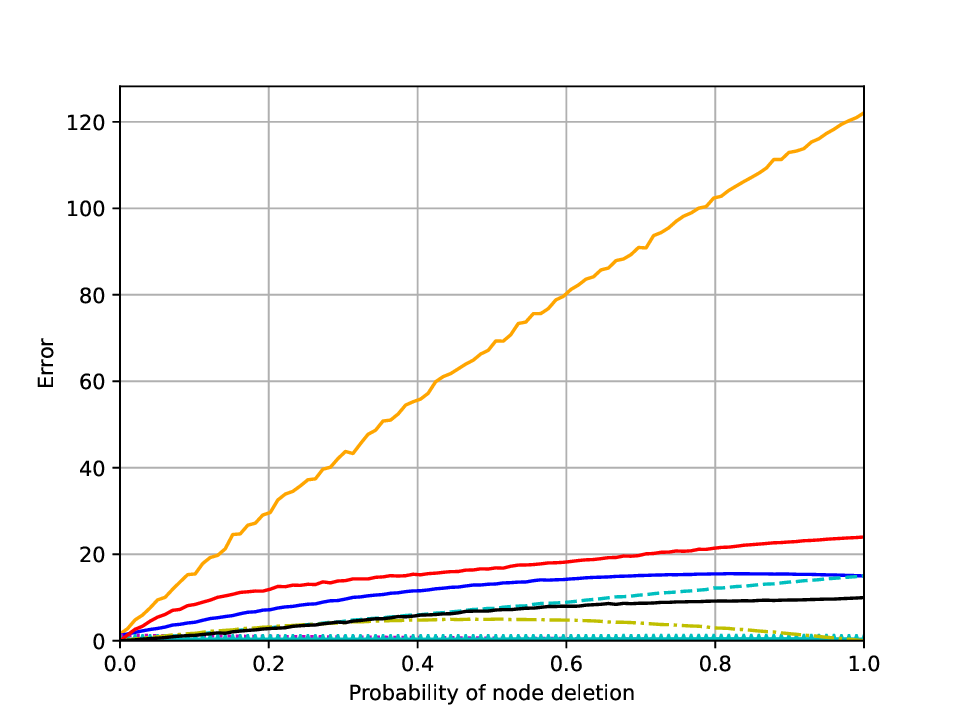}
  \caption{Random node removal.}
  \label{fig:exp_d}
\end{subfigure}
\caption{Plots of average graph GOSPA metric, GCD, GED and MCS  comparing the ground truth graph with (a) graphs with random Gaussian noise with increasing noise variance in node attributes; (b) graphs with random edge addition in adjacency matrix and Gaussian noise in node attributes; (c) graphs with random edge removal in adjacency matrix and Gaussian noise in node attributes; (d) graphs with different sizes and Gaussian noise in node attributes.}
\label{fig:exp}
\end{figure*}

\section{Experimental Results}\label{sec:experiments}

This section demonstrates the properties of the graph GOSPA metric by evaluating it on simulated undirected graphs (Section \ref{subsec:simulation undirected}) and simulated directed graphs (Section \ref{subsec:simulation directed}). We also demonstrate the application of the graph GOSPA metric on a real dataset with molecules (Section \ref{subsec:molecule dataset}). We compare the proposed metric with the MCS distance \cite[Thm. 1]{bunke1997relation}, GED \cite{fischer2015approximation} and GCD \cite[Eq. (2)]{Moharrer20}. The MCS distance between two graphs $X$ and $Y$ is $n_X+n_Y-2 n_{c}$, where $n_{c}$ represents the number of nodes in the largest common subgraph. In this section, we use the LP implementation of the graph GOSPA metric\footnote{Matlab and Python implementations of the graph GOSPA metric is available via Github: \url{https://github.com/JinhaoGu/The-graph-GOSPA-metric}.} and refer to it simply as graph GOSPA metric and the solver we use for LP is the interior-point method \cite{luenberger15}.


To use the GCD with graphs of different sizes, we use the procedure in \cite{Bento19} by adding dummy nodes to the smallest graph so that graphs have the same size. The dummy nodes have zero attribute value and are well-separated from any other nodes in the graphs. We have followed the publicly available code in \cite{Moharrer20} to write our own version of the GCD distance \cite[eq. (2)]{Moharrer20}, directly optimising over the doubly stochastic matrix that represents the node assignments. The MCS distance is computed by using the built-in function in RDkit to compute the MCS \cite{rdkit}. For the MCS distance, we do not use node attributes in the simulations, while in the dataset with molecules, we use the atom type as the node attribute. The GED is computed with the built-in function in NetworkX\cite{SciPyProceedings_11}. For the GED, we use a node match function such that, if the distance between two nodes is smaller than $c$, which is set to 3, they are treated as matched nodes and node deletion and insertion cost are set to 1. The edge substitution cost is set to 0 and the edge deletion and insertion cost are set to 1 as well. We set the hyper-parameters $p=1$, $\lambda=1$ for the GCD, and $p=1$, $c=3$, $\epsilon=1$ for the graph GOSPA metric. We also use the Euclidean distance as the base metric for both GCD and graph GOSPA metric. 

\subsection{Simulation results for undirected graphs}\label{subsec:simulation undirected}

In this section, we consider random graphs generated from a $G(n,p)$ Erdos-Renyi model \cite{erdHos1960evolution}, where $n$ represents the number of nodes in the graph, and $p$ represents the probability of an edge existing between two nodes. The nodes of the graphs have 2-D node states. These graphs can represent the locations of some users and the connectivities with other users. 

We generate a ground truth graph $X$ using the Erdos-Renyi model by setting $n=10$ and $p=0.4$ with the GraSP toolbox \cite{girault2017}, as shown in Figure \ref{fig:synthetic graph}, and the location elements of $X$ are generated by using the neato layout engine with parameter $maxiter=5000$ and $start=7$ in Graphviz toolkit \cite{ellson2002}. We then generate a modified version of this graph, denoted as $Y$, which varies across different scenarios. Each of these modified graphs $Y$ represents a different type of change or perturbation in the original graph $X$, such as adding noise to node attributes (Scenario 1), adding or removing edges (Scenario 2 and 3), or removing nodes (Scenario 4). By comparing $X$ and $Y$, we can assess how well the algorithm performs in detecting and characterizing these changes in the graph structure. 


We compute the average distances in each of the previous scenarios via Monte Carlo simulation with 1000 runs. In each Monte Carlo run for Scenario 1, we draw a graph $Y$ by adding independent zero-mean Gaussian random variables with covariance matrix $\sigma^2I_2$ to the attributes of $X$. The average errors for different values of $\sigma^2$ are shown in Figure \ref{fig:exp_a}. With increasing noise variance in node attributes, the GCD value keeps growing, which is caused by the growth of localisation error in GCD. In the graph GOSPA metric, when the localisation error is too large to assign two nodes to each other, the missed node error and false node error will increase until all nodes are missed/false nodes. Furthermore, the contribution of missed/false error is only determined by the quantity of missed/false nodes, therefore the graph GOSPA saturates, when there is a large noise variance in node attributes. For the GED, as the noise level increases, the error grows until it reaches its maximum, which is determined by the number of nodes in the corresponding graphs, which explains the saturation area in Figure \ref{fig:exp_a}.

In Scenario 2, we obtain graph $Y$ by adding edges in the adjacency matrix of ground truth $X$ with probability $q$ and we also add random noise with covariance matrix  $0.1 I_2$ to the node attributes. The average errors for different $q$ are shown in Figure \ref{fig:exp_b}. In Figure \ref{fig:exp_b}, the values of GED, GCD and graph GOSPA increase, where the error grows slowly as the probability of adding extra edges increases. We can also notice that the edge mismatch cost increases linearly with the increasing probability of adding edges, which plays a predominant role in determining the metric value of the graph using the GOSPA metric. For the MCS distance, as we are adding edges to the ground truth graph in Scenario 2, the maximum common subgraph between the target graph and the ground truth graph is actually the ground truth itself, therefore we get zero MCS distance error.

The graph $Y$ in Scenario 3 is generated by randomly deleting edges in ground truth $X$ with probability $q$, and we also add random noise with covariance matrix $0.1I_2$ to the node attributes. In Figure \ref{fig:exp_c}, we notice that the errors for all the considered methods are similar to the values in Figure \ref{fig:exp_b}, except for the MCS distance. This is due to the fact that the size of maximum common subgraphs is getting smaller with more edges being deleted from the prototype graph $X$.
Thus, the error increases in the MCS distance as the probability of edge deletion grows gradually. 

In Scenario 4, we draw graph $Y$ by randomly removing the nodes in the ground truth graph $X$ with a certain probability of node deletion. We also add Gaussian noise with covariance matrix $0.1I_2$ to node attributes. In Figure \ref{fig:exp_d}, we see that with the increasing probability of removing nodes, the number of unassigned nodes in graph GOSPA rises, missed nodes specifically, and the localisation error for assigned nodes decreases. The edge mismatch cost in this example grows first until the probability of node deletion $q=0.5$, then the edge mismatch cost decreases. This is because the graph GOSPA metric does not consider the edges connected to unassigned nodes. Thus, as there are more unassigned nodes with a higher probability of node deletion, the edge mismatch cost decreases. For the other graph distances, GCD, GED and the MCS distance, their error also increases with the probability of node deletion. For GCD, the error grows linearly, while for the other two distances the error grows similarly as in the graph GOSPA metric. 

\subsubsection*{Computational complexity evaluation}\label{subsubsec:computation analysis udirected}

In this section, we generate random undirected graphs using the same settings as in Section \ref{subsec:simulation undirected}, but we change the number of nodes in the generated graphs to $n_X=n_Y \in \{10,20,40,60,80,100\}$. Then, we calculate the time and memory usage of the implementations of the graph GOSPA metric for undirected graphs, GCD, GED and the MCS distance.

In Table \ref{tab:time compare undirected}, we can see that the graph GOSPA metric is the one with the lowest computational time due to the LP implementation. The second fastest is the GCD. The GED takes longer than 3000 seconds (used as a time limit) to compute even when we consider graphs with 20 nodes. The MCS also takes more than 3000 seconds to compute graphs with 20 nodes. The considered MCS implementation did not work for graphs with more than 40 nodes. The memory usage for each metric has been calculated using the \textit{tracemalloc} library in Python and is shown in Table \ref{tab:memory usage undirected}. The graph GOSPA metric requires more memory than GCD to store the constraints, but less memory than GED.

\begin{table}
    \centering
    \caption{Computational time over undirected graphs (CPU time in seconds).}
    \begin{tabular}{c|c|c|c|c}
    \hline 
       $(n_X,n_Y)$& Graph GOSPA& GCD & GED &MCS\\
    \hline 
    (10,10)  &0.02 & 0.03    & 58.26 & 5.81\\
    (20,20) & 0.06 & 0.20   & >3000 & >3000\\
    (40,40) & 0.97 & 8.78   & >3000 &-\\
    (60,60) & 6.71 & 102.70 & >3000 &-\\
    (80,80) & 45.29& 654.69 & >3000 &-\\
    (100,100)&131.40& 2550.10& >3000 &-\\
    \hline
    \end{tabular}
    \label{tab:time compare undirected}
\end{table}

\begin{table}
    \centering
    \caption{Memory usage over undirected graphs (in MBytes).}
    \begin{tabular}{c|c|c|c|c}
    \hline 
    $(n_X,n_Y)$& Graph GOSPA& GCD & GED&MCS \\
    \hline 
    (10,10) & 0.49   & 0.77   & 0.86 & 0.90\\
    (20,20) & 4.81   & 2.21   & 7.11& 4968.46\\
    (40,40) & 71.07  & 14.78  & 153.28&--\\
    (60,60) & 353.34 & 46.36  & 1084.15&--\\
    (80,80) &1111.18 & 108.27 & 4674.12&--\\
    (100,100)&2660.09 & 244.50 & 14272.59&--\\
    \hline
    \end{tabular}
    \label{tab:memory usage undirected}
\end{table}

\subsection{Simulation results for directed graphs}\label{subsec:simulation directed}
\begin{figure}[t]
  \centering
  \includegraphics[width=.8\linewidth]{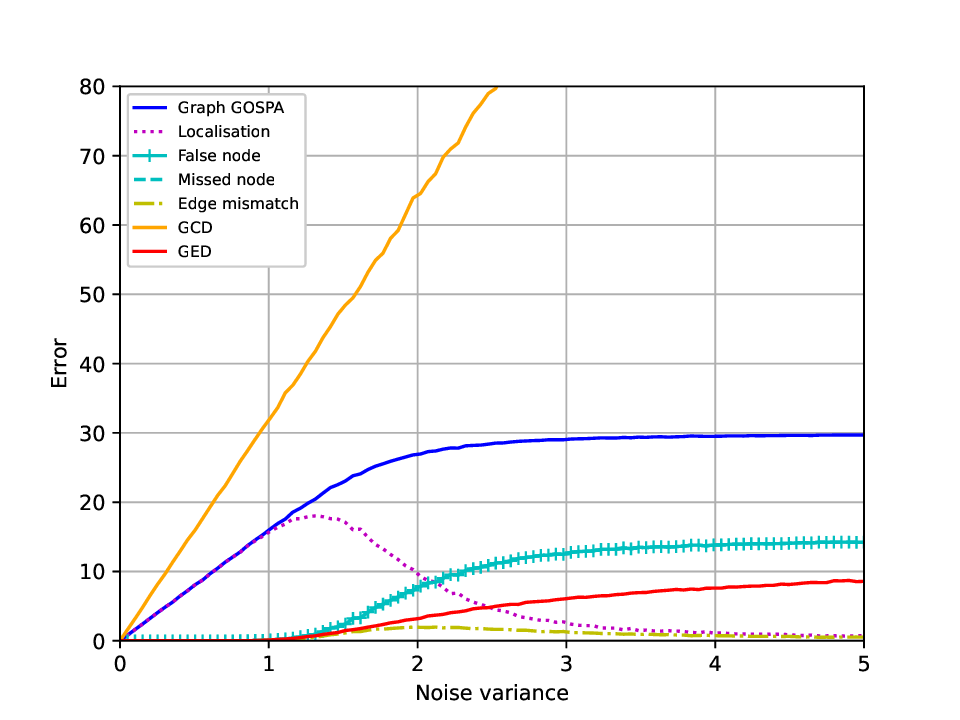}
\caption{Plots of average graph GOSPA metric, GCD and GED  comparing the directed ground truth graph with directed graphs with increasing random noise variance in node attributes.}
\label{fig:exp_di}
\end{figure}

In this section, we compare
the graph GOSPA metric with the GCD and the GED over directed graphs. In this comparison, we use the symmetrised GCD for directed graphs in \cite[Thm. 2]{Bento19}, which requires solving two optimisation problems. We consider the same graphs used in Scenario 1 in Section \ref{subsec:simulation undirected}, but setting zeros in the lower triangle of the adjacency matrix to make the graphs directed. The metric values for increasing noise variance $\sigma^2$ in the attributes are shown in Figure \ref{fig:exp_di}.

The graph GOSPA metric error in Figure \ref{fig:exp_di} is similar to the undirected case, only grows slower than the graph GOSPA error in Figure \ref{fig:exp_a}, since the edge penalties for edges connected to unassigned nodes are halved. The GCD in Figure \ref{fig:exp_di}, is also similar but twice the value due to how the extension to directed graphs is done in \cite[Thm. 2]{Bento19}. The GED remains the same as the undirected case, although we removed one direction in the compared graphs.


\subsubsection*{Computational complexity evaluation}\label{subsubsec:complexity directed}
In this section, we use the same graphs used for the computational analysis in Section \ref{subsubsec:computation analysis udirected}, but we replace the lower triangle of the adjacency matrices with zeros to obtain directed graphs. We also compute the time and memory usage for directed graphs.
The simulation results are summarised in Table \ref{tab:time compare directed} and \ref{tab:memory usage directed}. As with undirected graphs, the graph GOSPA metric is the one with the lowest computational complexity, though it has higher memory usage than the GCD.

\begin{table}
    \centering
    \caption{Computational time over directed graphs (CPU time in seconds).}
    \begin{tabular}{c|c|c|c}
    \hline 
       $(n_X,n_Y)$& Graph GOSPA& GCD & GED \\
             
    \hline 
    (10,10)  & 0.01  & 0.5    & 58.34\\
    (20,20)  & 0.05  & 0.37   & >3000\\
    (40,40)  & 0.93  & 16.44   & >3000\\
    (60,60)  & 6.37 & 203.28 & >3000\\
    (80,80)  & 26.41 & 1269.95 & >3000\\
    (100,100)& 84.24& 5119.86& >3000\\
    \hline
    \end{tabular}
    \label{tab:time compare directed}
\end{table}

\begin{table}
    \centering
    \caption{Memory usage over directed graphs (in MBytes).}
    \begin{tabular}{c|c|c|c}
    \hline 
    $(n_X,n_Y)$&  Graph GOSPA& GCD & GED \\
            
    \hline 
    (10,10) & 0.79    & 0.77   & 0.83\\
    (20,20) & 10.38   & 2.26   & 11.44\\
    (40,40) & 158.77  & 14.83  & 166.14\\
    (60,60) & 787.80  & 46.40  & 1107.38\\
    (80,80) & 2473.88 & 108.31 & 4703.80\\
    (100,100)& 6008.09 & 244.53 & 14315.83\\
    \hline
    \end{tabular}
    \label{tab:memory usage directed}
\end{table}

\subsection{Experiments on a molecular dataset}\label{subsec:molecule dataset}
\begin{figure}[t]
\begin{subfigure}{.5\textwidth}
  \centering
  \includegraphics[width=.9\linewidth]{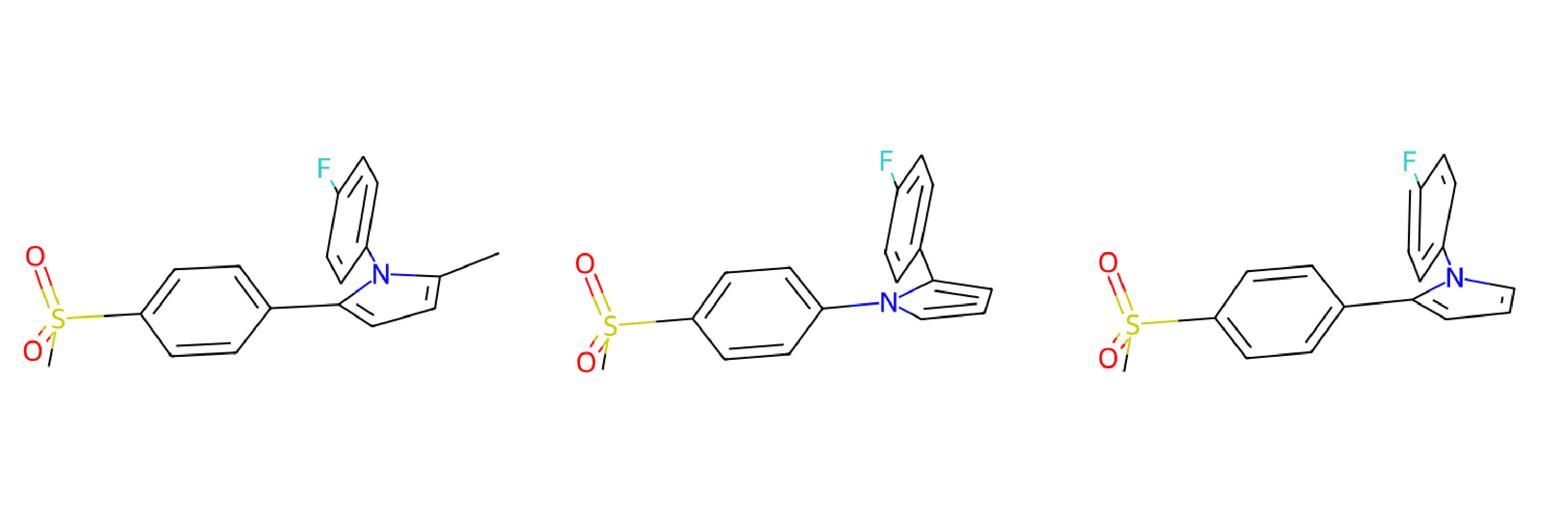}
  \caption{COX-2}
  \label{fig:cox}
\end{subfigure}
\begin{subfigure}{.5\textwidth}
  \centering
  \includegraphics[width=.9\linewidth]{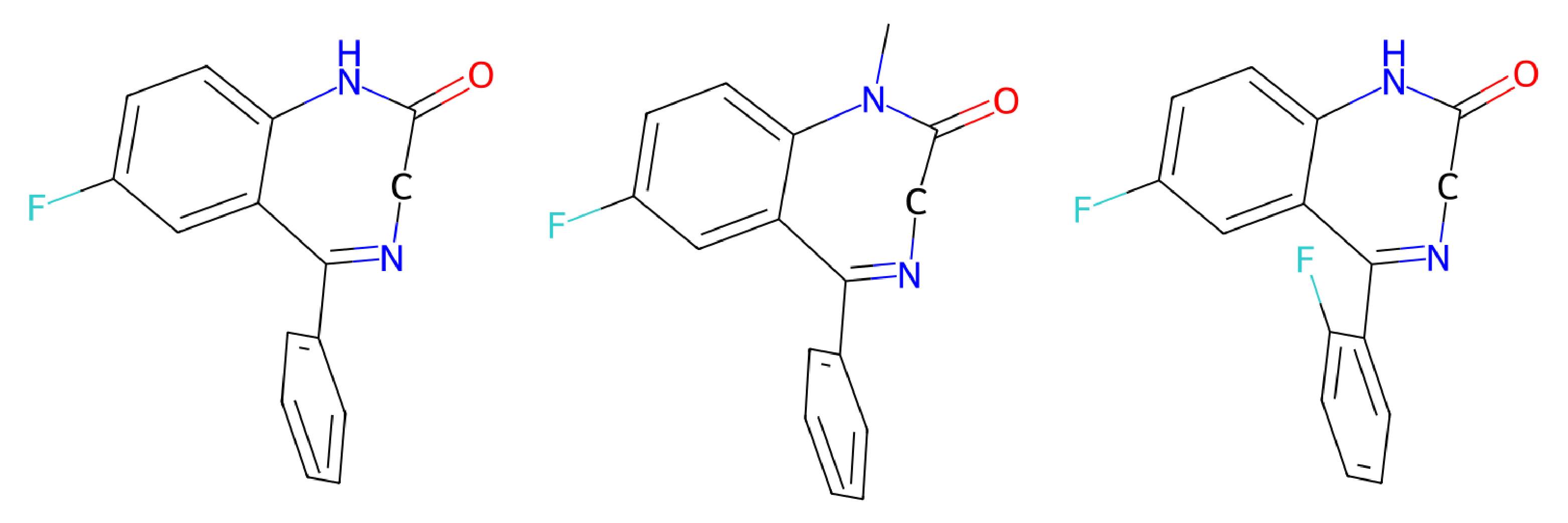}
  \caption{BZR}
  \label{fig:bzr}
\end{subfigure}
\begin{subfigure}{.5\textwidth}
  \centering
  \includegraphics[width=0.9\linewidth]{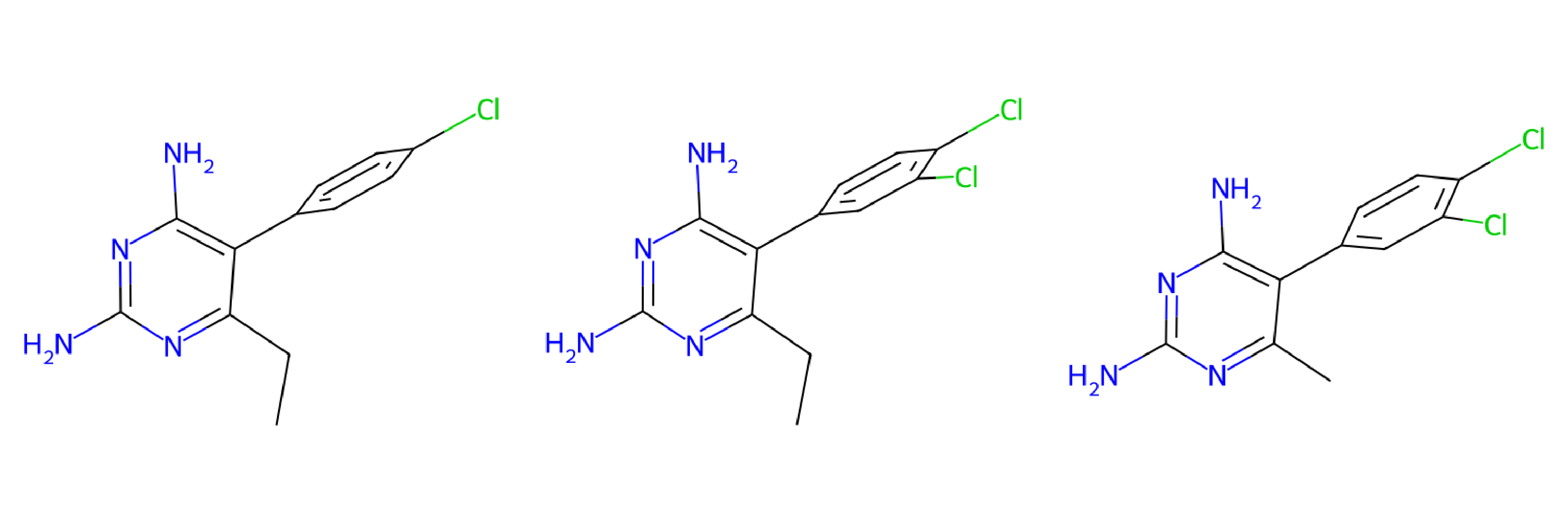}
  \caption{DHFR}
  \label{fig:dhfr}
\end{subfigure}
\caption{Plots of three molecular graph samples from the following collections: (a) COX-2; (b) BZR; (c) DHFR.}
\label{fig:mol}
\end{figure}

In this section, we illustrate the use of the graph GOSPA metric on the molecular dataset in the supplementary material of \cite{jeffrey2003}. This molecular dataset contains five collections of different types of compounds and we use three collections of them: cyclooxygenase-2 inhibitors (COX-2, 467 graphs), benzodiazepine receptor ligands (BZR, 405 graphs), dihydrofolate reductase inhibitors (DHFR, 756 graphs). All collections have molecules with more than 20 heavy atoms (C, N, O, Cl, etc.). For simplicity, we choose three molecules from each collection, and nine molecules in total. Within each collection, the molecular graphs have similar structures as those shown in Figure \ref{fig:mol}.

\begin{figure}[t]
\begin{subfigure}{.49\columnwidth}
  \centering
  \includegraphics[width=.9\linewidth]{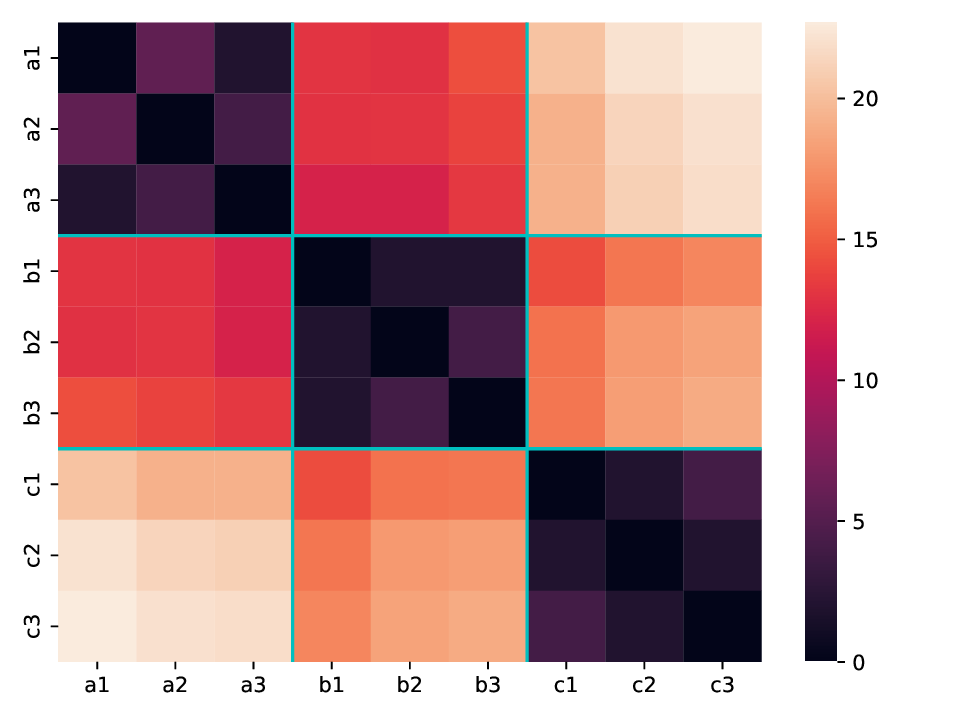}
  \caption{Graph GOSPA metric}
  \label{fig:mol gospa}
\end{subfigure}
\begin{subfigure}{.49\columnwidth}
  \centering
  \includegraphics[width=.9\linewidth]{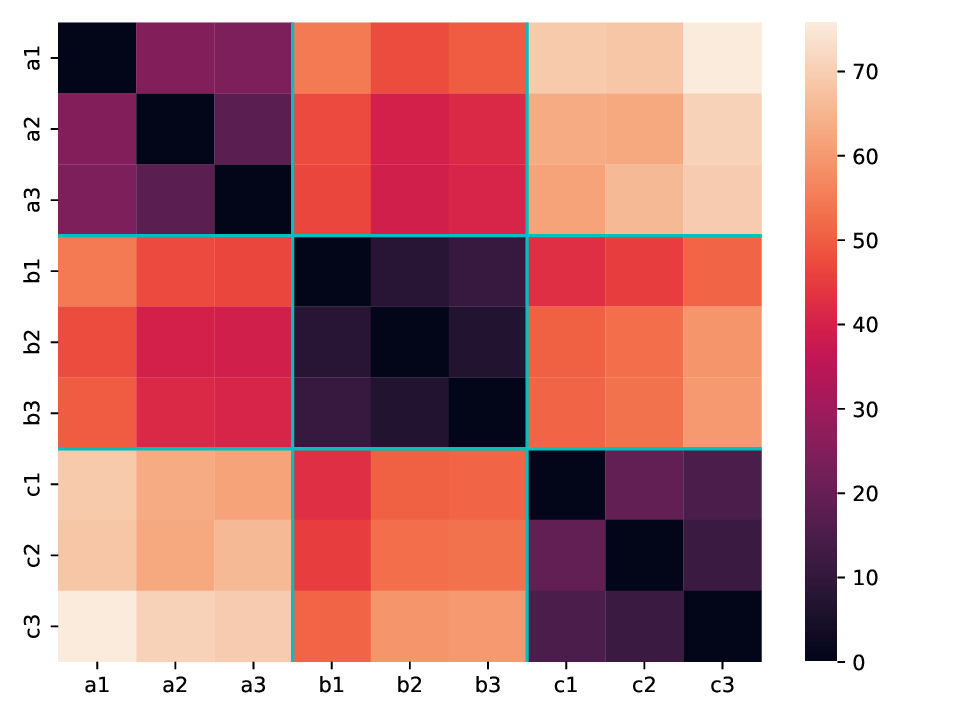}
  \caption{GCD}
  \label{fig:mol gcd}
\end{subfigure}

\begin{subfigure}{.49\columnwidth}
  \centering
  \includegraphics[width=.9\linewidth]{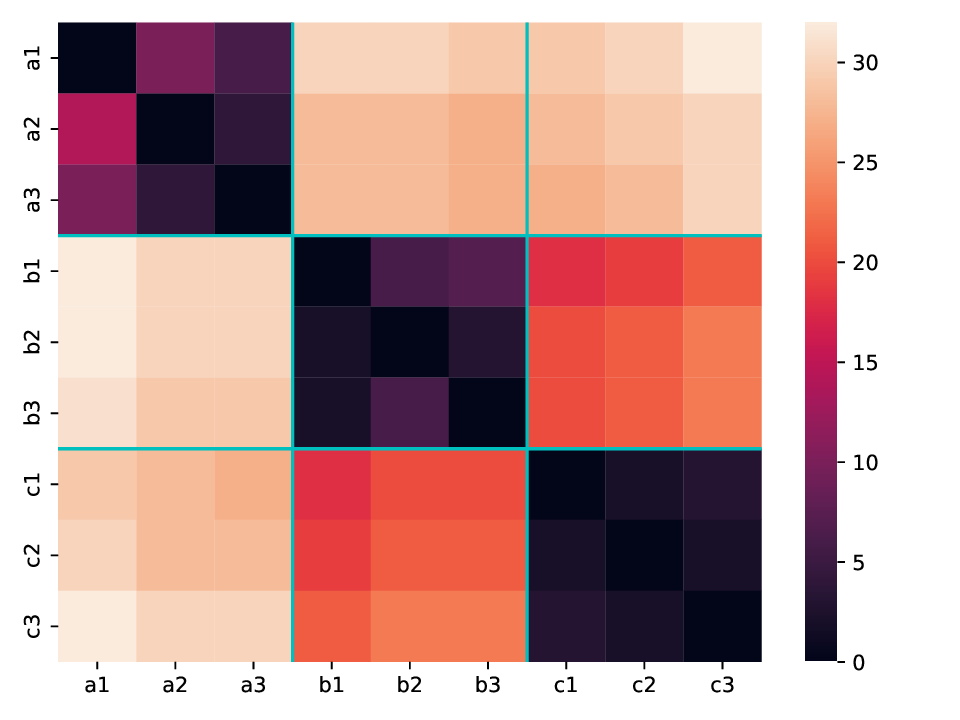}
  \caption{GED}
  \label{fig:mol ged}
\end{subfigure}
\begin{subfigure}{.49\columnwidth}
  \centering
  \includegraphics[width=.9\linewidth]{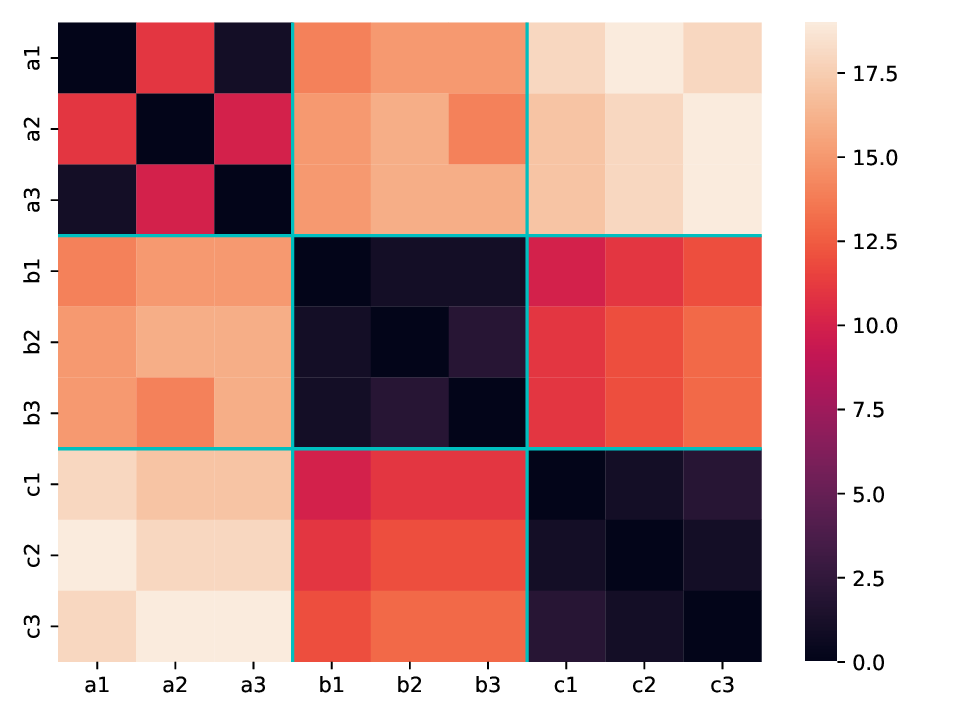}
  \caption{MCS distance}
  \label{fig:mol mcs}
\end{subfigure}

\caption{Illustration of the distance matrices computed using (a) graph GOSPA, (b) GED, (c) GED and (d) GCD for molecules from three different collections.
Each collection's metric value is separated by a solid line. Each entry of the $x$ and $y$ axes is named as `a$i$`, `b$i$` or `c$i$` where $i$ is either 1, 2 or 3, and represents the $i$-th molecule in the 'a', 'b' or 'c' family in Figure \ref{fig:mol}.}
\label{fig:distance matrix}
\end{figure}

In this experiment, we use the atomic number of each element (or node) in the molecular graphs as the node attributes. In addition, as there are different bonds within molecules, the molecular graphs are weighted graphs, in which the entries of the adjacency matrices can be $0$, $1$, $2$, to represent different types of bonds. That is, weight $0$ means there is no bond, weight $1$ represents a single bond, and weight $2$ represents a double bond. We set the hyperparameters $p=1$, $\lambda=1$ for the GCD distance, $p=1$, $c=3$, $\epsilon=1$ for the graph GOSPA metric. Both the GCD and the graph GOSPA metric use the Euclidean distance as the base metric.

In Figure \ref{fig:distance matrix}, we compare four different graph distances on molecules from three different types of collections. The results of graph GOSPA, GCD, GED and MCS distance are shown in Figure \ref{fig:mol gospa}, \ref{fig:mol gcd}, \ref{fig:mol ged} and \ref{fig:mol mcs} respectively. In all four distance matrices, different types of molecules are separated by a solid line. This implies that, in the diagonal blocks, we are comparing the molecules from the same family, whose distances should in principle be lower than the distances of molecules between different families. All distances manage to keep the property, though the MCS distance has more problems with the COX-2 family.

We now proceed to compare the molecule COX-2,1-1, considered as the ground truth graph $X$, with molecules from other collections. We use 100 molecules from each collection and compute the average distance value for every collection. In Table \ref{tab:mol0 average distance}, we show that, in graph GOSPA, the average distance between COX-2,1-1 and molecules in COX-2, is lower than the average distances with molecules in BZR and DHFR, as the structure becomes more diverse to the reference molecule COX-2,1-1. In this experiment, all considered distances work well, though the GCD performs a bit worse for the molecules in family COX-2. 

A feature of the graph GOSPA metric is that we can analyse the results using the metric decomposition. To this end, in Table \ref{tab:mol1 average distance decomposition}, we show the decomposition of the graph GOSPA in the previous experiment that considered molecule COX-2,1-1 as the ground truth graph $X$. 
In the experiment, we use the atom number as the node attribute, the localisation error can indicate the difference between atom types of two molecules. For example, suppose two molecules have similar structures, and almost all nodes within two molecular graphs have been assigned. In that case, the localisation error will show the difference in the atom types between two assigned nodes. The false node error shows the part that exists in the compared molecule but does not exist in the ground truth molecule. Similarly, the miss node error tells us some substructures exist in the ground truth but do not exist in the compared molecule. As for the edge mismatch cost, we can see the difference in molecule structures. The difference in bonds types, whether there is a bond between two assigned atoms (nodes) or if the bond type is different, they all contribute to the edge mismatch cost.

In COX-2 collection, the false node error and edge mismatch error are predominant in the metric value, which indicates that the other molecules in this family have substructures that do not exist in the ground truth molecule COX-2,1-1.
In BZR and DHFR, the metric value is mostly affected by the missed node and edge mismatch cost, which suggests that there are differences in the chemical structure.

\begin{table}
    \centering
    \caption{Average distance between molecule COX-2,1-1 and 100 molecules from each family.}
    \begin{tabular}{c|c|c|c}
    \hline 
        & COX-2 & BZR & DHFR \\
    \hline 
    Graph GOSPA  &11.78 & 19.15 & 23.49  \\
    GCD & 51.58 & 52.17 & 70.27 \\
    GED & 21.42 & 32.57 & 37.65 \\
    MCS & 15.54 & 17.41 & 23.15 \\
   
    \hline
    \end{tabular}
    \label{tab:mol0 average distance}
\end{table}

\begin{table}
    \centering
    \caption{Decomposition of the average graph GOSPA error between molecule COX-2,1-1 and 100 molecules from each family.}
    \begin{tabular}{c|c|c|c}
    \hline 
        & COX-2 & BZR & DHFR \\
    \hline 
    Localisation&2.04 & 3.36 & 4.75  \\
    False node& 4.94 & 3.45 & 6.74 \\
    Miss node& 0.24 & 4.16 & 3.70 \\
    Edge mismatch& 4.56 & 8.19 & 8.30 \\
   
    \hline
    \end{tabular}
    \label{tab:mol1 average distance decomposition}
\end{table}

\section{Conclusion}\label{sec:conclusion}
In this paper, we have proposed the graph GOSPA metric, which is based on obtaining the optimal assignment matrix between nodes, to measure the discrepancy between two graphs of different sizes. The graph GOSPA metric meets the metric properties (identity, symmetry and triangle inequality) for graphs with attributes and different sizes. The proposed metric has an intuitive decomposition that captures the node attribute errors for properly assigned nodes, missed and false node costs, and edge mismatch costs. 

For metric computation, we have proposed a lower bound to the graph GOSPA metric that is also a metric and can be computed using linear programming with polynomial time complexity. We have also extended this metric to weighted and directed graphs. 

A possible line of future work is the distributed implementation of the metric using for example the alternating direction of multipliers methods (ADMM) \cite{Moharrer20,boyd2011distributed} or other distributed optimisation methods \cite{jiang17,chellapandi23}. We are also going to explore the use of the graph GOSPA metric in different applications such as network science and chemistry in future work.

\bibliographystyle{IEEEtran}
\bibliography{references}

\clearpage
\begin{center}
	{\LARGE{}Supplementary material of "Graph GOSPA metric: a metric to measure the discrepancy between graphs of different sizes"}
	\par\end{center}{\LARGE \par}
\appendices

\section{}\label{sec:proof of lemma 1}
In this appendix, we prove that \eqref{eq:graph_metric} can be written as in \eqref{eq:simplified_graph_metric} in Lemma \ref{lem:graph_gospa}. We know from \cite{Rahmathullah17} that the localisation error, missed and false node costs can be written in matrix form as $\mathrm{tr}\big[D_{X,Y}^{\top}W\big]$. We proceed to prove the edge mismatch penalty \eqref{eq:edge_cost_mat}. We know that the assignment vector can be written as a binary matrix meeting \eqref{eq:binary_constraint1}-\eqref{eq:binary_constraint4}.
We note that the following sum in \eqref{eq:edge_cost_no_epsilon} can be written as 
\begin{flalign}\label{eq:A_xW}
    \sum_{k=1}^{n_{X}}A_{X}(i,k)W(k,j) & =\begin{cases}
A_{X}(i,k_{x}) & 
\exists k_x : W(k_x,j)=1, 
\\
0 & W(n_X+1,j)=1
\end{cases}
\end{flalign}
where the first line represents that node $k_x$ in $X$ is assigned to node $j$ in $Y$, and the second line means node $j$ in $Y$ is unassigned.
Equivalently, the following sum in (16) can be written as
\begin{flalign}\label{eq:WA_y}
    \sum_{k=1}^{n_{Y}}W(i,k)A_{Y}(k,j) & =\begin{cases}
A_{Y}(k_{y},j) & \exists k_y : W(i,k_y)=1,
\\
0 & W(i,n_Y+1)=1
\end{cases}
\end{flalign}
where the first line represents that node $k_y$ in $Y$ is assigned to node $i$ in $X$, and the second line means node $i$ in $X$ is unassigned. 

In \eqref{eq:edge_cost_no_epsilon} we then go through each node $i$ in $X$ and each node $j$ in $Y$. For the node $j$ in $Y$, we find the node $k_x$ in $X$ assigned to it (if it exists), using \eqref{eq:A_xW}.  For the node $i$ in $X$, we find the node $k_y$ in $Y$ assigned to it (if it exists), using \eqref{eq:WA_y}. If both $i$ and $j$ are properly assigned, there is an edge penalty if one of the graphs has the corresponding edge but the other graph does not. That is, either there is an edge $\{i,k_x\}$ in $X$ or $\{j,k_y\}$ in $Y$, but not both simultaneously. These types of edges are penalised twice in the above sum, as we will obtain the same result when $i=k_x$ and $j=k_y$ i.e., we are evaluating the node on the other side of the edge. This corresponds to the first line of \eqref{eq:edge_cost}.

If node $j$ in $Y$ is unassigned, there is an edge penalty if node $i$ is assigned to $k_y$ and there is an edge linking $k_y$ to $j$ in $Y$. That is, each edge in which one of the ends is assigned but the other end is unassigned, contributes to a half-edge error (as these errors are not repeated in the above sum).
This corresponds to the third line of \eqref{eq:edge_cost}.

For the case that node $i$ in $X$ is unassigned, it is identical to the previous case that node $j$ in $Y$ is unassigned, since \eqref{eq:edge_cost_no_epsilon} is symmetric. This corresponds to the second line in \eqref{eq:edge_cost}.
Therefore, we have proved that \eqref{eq:graph_metric} can be written in the form of \eqref{eq:simplified_graph_metric}. 

\section{}\label{sec:proof of triangle inequality}
In this appendix, we prove the triangle inequality property of
the graph metric in \eqref{eq:LP graph metric} for both undirected and directed graphs, as the identity
and symmetry properties of the metric are immediate from the definition. The proof in this section follows the proof in \cite{Angel20_d},
and we use the same notation as well.
\subsection{Proof of the triangle inequality property for undirected graphs}\label{proof:triangle undirected}
We denote the objective function in (\ref{eq:LP graph metric}),
$\bar{d}^{\left(c,\epsilon\right)}\left(X,Y,\mathrm{\mathit{W}}\right)$, as
a function of the assignment matrix $W$. The proof of the triangle
inequality is as follows. 

Let $\mathcal{\mathit{W}_{\mathit{X,Y}}^{*}\in}\mathcal{\overline{W}}_{X,Y}$, $\mathit{W}_{\mathit{X,Z}}^{*}\in\overline{\mathcal{W}}_{\mathit{X,Z}}$, $\mathit{W}_{\mathit{Z,Y}}^{*}\in\overline{\mathcal{W}}_{Z,Y}$
be the weight matrices which minimise $\bar{d}^{\left(c,\epsilon\right)}\left(X,Y,W_{X,Y}\right)$, 
$\bar{d}^{\left(c,\epsilon\right)}\left(X,Z,W_{X,Z}\right)$ and $\bar{d}^{\left(c,\epsilon\right)}\left(Z,Y,W_{Z,Y}\right)$, 
respectively. Given the matrices $\mathit{W}_{\mathit{X,Z}}^{*}$ and $\mathit{W}_{\mathit{Z,Y}}^{*}$, we define a matrix $\mathit{W}_{\mathit{X,Y}}$ that meets
\begin{equation}
\begin{aligned}
&W_{X,Y}(i,j)\\
&=\begin{cases}
1-\sum_{j=1}^{n_{\mathbf{\mathit{Y}}}}W_{\mathbf{\mathbf{\mathrm{\mathit{X}}}\mathit{\mathrm{,}Y}}}(i,j) & j=n_{Y}+1,i=1,\ldots,n_{X}\\
1-\sum_{i=1}^{n_{\mathbf{\mathit{X}}}}W_{\mathbf{\mathit{X}},\mathbf{\mathit{Y}}}(i,j) & i=n_{X}+1,j=1,\ldots,n_{Y}\\
0 & i=n_{X}+1,j=n_{Y}+1\\
\sum_{l=1}^{n_{\mathbf{\mathit{Z}}}}W_{\mathit{\mathit{X}},Z}^{*}(i,l)W_{Z,Y}^{*}(l,j) & \mathrm{otherwise.}
\end{cases}\label{eq:weight_matrices}
\end{aligned}
\end{equation}

Then we show that
\begin{equation}
\bar{d}^{\left(c,\epsilon\right)}\left(X,Y,W_{X,Y}\right)\leq \bar{d}^{\left(c,\epsilon\right)}\left(X,Z\right)+\bar{d}^{\left(c,\epsilon\right)}\left(Z,Y\right).\label{eq:triangle_inequality}
\end{equation}
It is met that $\bar{d}^{\left(c,\epsilon\right)}\left(X,Y\right)\leq \bar{d}^{\left(c,\epsilon\right)}\left(X,Y,W_{X,Y}\right)$, therefore, \eqref{eq:triangle_inequality} implies that the triangle inequality holds
\begin{flalign}
\bar{d}^{\left(c,\epsilon\right)}\left(X,Y\right)\leq \bar{d}^{\left(c,\epsilon\right)}\left(X,Z\right)+\bar{d}^{\left(c,\epsilon\right)}\left(Z,Y\right)
\end{flalign}
More generally, we show that for any $\mathit{\mathcal{\mathit{W}_{\mathrm{\mathit{X,Y}}}\in\overline{W}_{\mathit{X,Y}}},\mathit{W}_{\mathrm{\mathit{X,Z}}}\in\overline{\mathcal{W}}_{\mathrm{\mathit{X,Z}}},W_{\mathrm{\mathit{Z,Y}}}\in\overline{\mathcal{W}}_{\mathit{\mathrm{\mathit{Z,Y}}}}}$
generated from (\ref{eq:weight_matrices}), the following inequality
holds:
\begin{flalign}
\bar{d}^{\left(c,\epsilon\right)}&\left(X,Y,W_{X,Y}\right)\nonumber
\\&\leq \bar{d}^{\left(c,\epsilon\right)}\left(X,Z,W_{X,Z}\right)+\bar{d}^{\left(c,\epsilon\right)}\left(Z,Y,W_{Z,Y}\right).\label{eq:triangle_inequality1}
\end{flalign}
To show that \eqref{eq:triangle_inequality1} holds, we first obtain inequalities for the terms corresponding to $\mathrm{tr}\big[D_{X,Y}^{\top}W\big]$ and to the edge mismatch costs separately. The inequality for $\mathrm{tr}\big[D_{X,Y}^{\top}W\big]$ is given by (46) in \cite{Angel20_d}. In the next subsection, we will prove the inequality of the edge mismatch: $e_{X,Y}(W_{X,Y})^{p} \leq e_{X,Z}(W_{X,Z})^{p}+e_{Z,Y}(W_{Z,Y})^{p}$. 

Based on these two inequalities, the triangle inequality proof is finalised by using the Minkowski inequality. The steps are similar to those from Eq. (48) to (52) in \cite{Angel20_d} and are not included here for brevity. Therefore, we just need to prove the following edge mismatch cost inequality.

\subsubsection{Edge mismatch cost inequality}
For the edge mismatch cost $e_{X,Y}(W_{X,Y})$, we show that
\begin{align}
e_{X,Y}^{(\epsilon=2,p=1)}(W_{X,Y})=&\sum_{i=1}^{n_{X}}\sum_{j=1}^{n_{Y}}\left|\sum_{k=1}^{n_{X}}A_{X}(i,k)W_{X,Y}(k,j) \right. \nonumber \\
&-\left.\sum_{k=1}^{n_{Y}}W_{X,Y}(i,k)A_{Y}(k,j)\right|\nonumber \\
  \leq&\sum_{i=1}^{n_{X}}\sum_{l=1}^{n_{Z}}\left|\sum_{k=1}^{n_{X}}A_{X}(i,k)W_{X,Z}(k,l) \right. \nonumber \\
  &-\left. \sum_{k=1}^{n_{Z}}W_{X,Z}(i,k)A_{Z}(k,l)\right|\nonumber \\
 &+\sum_{j=1}^{n_{Y}}\sum_{l=1}^{n_{Z}}\left|\sum_{k=1}^{n_{Z}}A_{Z}(l,k)W_{Z,Y}(k,j) \right. \nonumber \\
 &-\left. \sum_{k=1}^{n_{Y}}W_{Z,Y}(l,k)A_{Y}(k,j)\right|\label{eq:edge_mismatch_triangle_inequality}.
\end{align}
where we have used $\epsilon=2$ and $p=1$ without loss of generality, as $\epsilon$ and $p$ are just part of the multiplication constant in \eqref{eq:edge_cost_no_epsilon}. 

From (\ref{eq:weight_matrices}), we know that $W_{X,Y}(i,j)=\sum_{l=1}^{n_{Z}}W_{X,Z}(i,l)W_{Z,Y}(l,j)$,
when $i\leq n_{X},j\leq n_{Y}$, thus, the left hand side of (\ref{eq:edge_mismatch_triangle_inequality})
becomes
\begin{flalign}
  \sum_{i=1}^{n_{X}}\sum_{j=1}^{n_{Y}}&\left|\sum_{k=1}^{n_{X}}\sum_{l=1}^{n_{Z}}A_{X}(i,k)W_{X,Z}(k,l)W_{Z,Y}(l,j)\right. \nonumber\\
 &-\left.\sum_{k=1}^{n_{Y}}\sum_{l=1}^{n_{Z}}W_{X,Z}(i,l)W_{Z,Y}(l,k)A_{Y}(k,j)\right|.
\end{flalign}
Here, we add and subtract the same term to the previous equation, and also drop the dependency of $e_{X,Y}$ on $W_{X,Y}$ for notational clarity
\begin{flalign}\label{eq:subtraction}
e_{X,Y}^{(2,1)}=  \sum_{i=1}^{n_{X}}\sum_{j=1}^{n_{Y}}&\left|\sum_{k=1}^{n_{X}}\sum_{l=1}^{n_{Z}}A_{X}(i,k)W_{X,Z}(k,l)W_{Z,Y}(l,j) \right. \nonumber \\
-&\left.\sum_{l=1}^{n_{Z}}\sum_{k=1}^{n_{Z}}W_{X,Z}(i,k)A_{Z}(k,l)W_{Z,Y}(l,j) \right. \nonumber \\
 +& \left.\sum_{l=1}^{n_{Z}}\sum_{k=1}^{n_{Z}}W_{X,Z}(i,k)A_{Z}(k,l)W_{Z,Y}(l,j) \right. \nonumber\\ 
 -&\left.\sum_{k=1}^{n_{Y}}\sum_{l=1}^{n_{Z}}W_{X,Z}(i,l)W_{Z,Y}(l,k)A_{Y}(k,j)\right|
`\end{flalign}

\begin{flalign}\label{eq:swap}
=  \sum_{i=1}^{n_{X}}\sum_{j=1}^{n_{Y}}&\left|\sum_{l=1}^{n_{Z}}\left(\sum_{k=1}^{n_{X}}A_{X}(i,k)W_{X,Z}(k,l)
\right. \right. \nonumber \\ 
-&\left. \left. \sum_{k=1}^{n_{Z}}W_{X,Z}(i,k)A_{Z}(k,l)\right)W_{Z,Y}(l,j) \right. \nonumber \\
 +& \left.\sum_{l=1}^{n_{Z}}W_{X,Z}(i,l)\left(\sum_{k=1}^{n_{Z}}A_{Z}(l,k)W_{Z,Y}(k,j)\right. \right. \nonumber\\
 -&\left. \left. \sum_{k=1}^{n_{Y}}W_{Z,Y}(l,k)A_{Y}(k,j)\right) \right|.
\end{flalign}

Note that the range of $k$ and $l$ in \eqref{eq:subtraction} above are the same, so they can be swapped in \eqref{eq:swap}. Then, we apply the inequality $\left|\sum_{l}a_{l}\right|\leq\sum_{l}\left|a_{l}\right|$
to the equation above, yields
\begin{flalign}
e_{X,Y}^{(2,1)} \leq & \sum_{i=1}^{n_{X}}\sum_{j=1}^{n_{Y}}\sum_{l=1}^{n_{Z}}
\bigg|\left(\sum_{k=1}^{n_{X}}A_{X}(i,k)W_{X,Z}(k,l) \right.
     \nonumber \\
    &\left. -\sum_{k=1}^{n_{Z}}W_{X,Z}(i,k)A_{Z}(k,l) \right) W_{Z,Y}(l,j)  \nonumber \\
     & + W_{X,Z}(i,l)\left(\sum_{k=1}^{n_{Z}}A_{Z}(l,k)W_{Z,Y}(k,j)\right. \nonumber\\
     &\left. -\sum_{k=1}^{n_{Y}}W_{Z,Y}(l,k)A_{Y}(k,j) \right) 
     \bigg| \\
\leq & \sum_{i=1}^{n_{X}}\sum_{l=1}^{n_{Z}}\sum_{j=1}^{n_{Y}} \bigg|\left(\sum_{k=1}^{n_{X}}A_{X}(i,k)W_{X,Z}(k,l) \right. \nonumber \\
 &\left.-\sum_{k=1}^{n_{Z}}W_{X,Z}(i,k)A_{Z}(k,l)\right)W_{Z,Y}(l,j)\bigg|  \nonumber\\
 & + \sum_{j=1}^{n_{Y}}\sum_{l=1}^{n_{Z}}\sum_{i=1}^{n_{X}} \nonumber \\ 
 &\bigg|W_{X,Z}(i,l)\left(\sum_{k=1}^{n_{Z}}A_{Z}(l,k)W_{Z,Y}(k,j) \right. \nonumber \\
 &\left.-\sum_{k=1}^{n_{Y}}W_{Z,Y}(l,k)A_{Y}(k,j)\right)\bigg|.\label{eq:absolute_value}
\end{flalign}

We now use the absolute value of $|ab|=|a||b|$ for two real numbers $a$ and $b$. From \eqref{eq:binary_constraint5}, we know that the weight matrices $W$ are non-negative. Note that the sum over $j$ only affects weight matrix $W_{Z,Y}$ and $i$ only affects weight matrix $W_{X,Z}$, therefore, based on \eqref{eq:binary_constraint1}, we can write \eqref{eq:absolute_value} as 
\begin{flalign}
    \sum_{i=1}^{n_{X}}&\sum_{l=1}^{n_{Z}}\sum_{j=1}^{n_{Y}} \bigg|\sum_{k=1}^{n_{X}}A_{X}(i,k)W_{X,Z}(k,l)  \nonumber \\
 -&\sum_{k=1}^{n_{Z}}W_{X,Z}(i,k)A_{Z}(k,l)\bigg| W_{Z,Y}(l,j)  \nonumber\\
 +&  \sum_{j=1}^{n_{Y}}\sum_{l=1}^{n_{Z}}\sum_{i=1}^{n_{X}} \nonumber \\ 
 &W_{X,Z}(i,l)\bigg|\sum_{k=1}^{n_{Z}}A_{Z}(l,k)W_{Z,Y}(k,j)  \nonumber \\
 &-\sum_{k=1}^{n_{Y}}W_{Z,Y}(l,k)A_{Y}(k,j)\bigg|.
\end{flalign}       
Thus, we can write the equation above as
\begin{flalign}
e^{(2,1)}_{X,Y}(W_{X,Y})\leq & \sum_{i=1}^{n_{X}}\sum_{l=1}^{n_{Z}}\bigg|\sum_{k=1}^{n_{X}}A_{X}(i,k)W_{X,Z}(k,l) \nonumber \\ 
&-\sum_{k=1}^{n_{Z}}W_{X,Z}(i,k)A_{Z}(k,l)\bigg| \nonumber \\
& +\sum_{j=1}^{n_{Y}}\sum_{l=1}^{n_{Z}}\bigg|\sum_{k=1}^{n_{Z}}A_{Z}(l,k)W_{Z,Y}(k,j)\nonumber \\
& -\sum_{k=1}^{n_{Y}}W_{Z,Y}(l,k)A_{Y}(k,j)\bigg|\\
= & e_{X,Z}^{(2,1)}(W_{X,Z})+e_{Z,Y}^{(2,1)}(W_{Z,Y})
\end{flalign}
and this proves (\ref{eq:edge_mismatch_triangle_inequality}). As explained before, this result, together with the derivations in \cite{Angel20_d}, proves \eqref{eq:triangle_inequality1}, which in turn proves Proposition \ref{prop:graphBinlp}. This also proves that \eqref{eq:graph_metric} is a metric, as the proof remains unchanged if we use the constraint \eqref{eq:binary_constraint4}.

\subsection{Proof of the triangle inequality property for directed graphs}\label{sec:triangle directed}
In this section, we prove the triangle inequality of the metric for directed graphs, see Lemma \ref{lem:extension directed graphs}. The proof of the triangle inequality is analogous to the proof for undirected graphs, which was provided in Appendix \ref{proof:triangle undirected}, except for the inequality in the edge mismatch cost. Therefore, in this appendix, we prove the required inequality for the edge mismatch cost.

The edge mismatch cost for directed graphs in \eqref{eq:edge cost directed} can be divided into two parts,
\begin{flalign}
    e_{X,Y}(W)=\frac{1}{2}\bigg(\tilde{e}_{X,Y}(W) +\tilde{e}^{(2)}_{Y,X}(W)\bigg)
\end{flalign}
where 
\begin{flalign}
    \tilde{e}_{X,Y}=\frac{\epsilon^p}{2} ||A_X W_{1:n_X,1:n_Y}- W_{1:n_X,1:n_Y}A_Y||
\end{flalign}
and 
\begin{flalign}
    \tilde{e}^{(2)}_{Y,X}&=\frac{\epsilon^p}{2}||A_Y W^T_{1:n_X,1:n_Y}-W^T_{1:n_X,1:n_Y}A_X||.
\end{flalign}
We have already proved in Appendix \ref{proof:triangle undirected} that, for any graphs $X$, $Y$, $Z$, $\tilde{e}_{X,Y} \leq \tilde{e}_{X,Z}+\tilde{e}_{X,Y}$. Importantly, to obtain this result, the adjacency matrices can have a general form (they are not required to be symmetric). Therefore, we can analogously
derive the inequality $\tilde{e}^{(2)}_{Y,X} \leq \tilde{e}^{(2)}_{Z,X}+\tilde{e}^{(2)}_{Y,Z}$. 
Therefore, we have
\begin{flalign}
    e_{X,Y}&=\frac{1}{2}\left(\tilde{e}_{X,Y}+ \tilde{e}^{(2)}_{Y,X}\right) \nonumber\\
           & \leq \frac{1}{2}\left( \tilde{e}_{X,Z}+\tilde{e}_{Z,Y}) + \tilde{e}^{(2)}_{Z,X}+\tilde{e}^{(2)}_{Y,Z}\right) \nonumber\\
           & =\frac{1}{2}(\tilde{e}_{X,Z}+ \tilde{e}^{(2)}_{Z,X})+ \frac{1}{2}(\tilde{e}_{Z,Y}+\tilde{e}^{(2)}_{Y,Z}) \nonumber \\
           & = e_{X,Z}+e_{Z,Y}.
\end{flalign}
Then, we can continue the proof as in the undirected case.

\subsection{Proof for computability using LP}\label{proof:LP proof}

The proof for the computability of the metric in \eqref{eq:LP graph metric} using LP follows the proof in \cite{Angel20_d}, \cite{Bento_draft16}. We note that the computation of the metric in \eqref{eq:LP graph metric} requires solving the following optimisation problem:
\begin{flalign}
    \underset{\substack{W\in\hW_{X,Y}}}{\arg\min}\left(\mathrm{tr}\big[D_{X,Y}^{\top}W\big]+e_{X,Y}(W)^p\right)
\end{flalign}
where $e_{X,Y}(W)$ is given in \eqref{eq:edge_cost_mat}.

\subsubsection{Proof for undirected graphs} \label{proof: undirected LP}

The objective function for undirected graphs can be written as 
\begin{flalign}\label{eq:objective function}
    \underset{\substack{W\in\hW_{X,Y}},\,  \hat{e}\in \mathbb{R}}{\arg\min}\left(\mathrm{tr}\big[D_{X,Y}^{\top}W\big]+\frac{\epsilon^p}{2} \hat{e}\right)
\end{flalign}
where we have introduced an extra variable $\hat{e} \in \R$ to the optimisation problem with constraint
\begin{flalign}\label{eq:constraint nonlinear}
    \hat{e} \geq \sum_{i=1}^{n_{X}}\sum_{j=1}^{n_{Y}}\left|\sum_{k=1}^{n_{X}}A_{X}(i,k)W(k,j) -\sum_{k=1}^{n_{Y}}W(i,k)A_{Y}(k,j)\right|,
\end{flalign}
All the constraints in \eqref{eq:binary_constraint1},\eqref{eq:binary_constraint2}, \eqref{eq:binary_constraint3} and \eqref{eq:binary_constraint5} are linear except the constraints in \eqref{eq:constraint nonlinear}. To make the optimisation problem linear, we can introduce additional variables $H(i,j) \in \R$ with the constraints:
\begin{flalign}
    \hat{e} &\geq \sum_{i=1}^{n_{X}}\sum_{j=1}^{n_{Y}} H(i,j), \label{eq:e>H}\\
    H(i,j) &\geq \sum_{k=1}^{n_{X}}A_{X}(i,k)W(k,j) -\sum_{k=1}^{n_{Y}}W(i,k)A_{Y}(k,j), \label{eq:h>abs}\\
    H(i,j) &\geq \sum_{k=1}^{n_{Y}}W(i,k)A_{Y}(k,j) -\sum_{k=1}^{n_{X}}A_{X}(i,k)W(k,j). \label{eq:h>-abs}
\end{flalign}

Now, the optimisation problem \eqref{eq:objective function} is linear with linear constraints \eqref{eq:binary_constraint1}, \eqref{eq:binary_constraint2}, \eqref{eq:binary_constraint3}, \eqref{eq:binary_constraint4}, \eqref{eq:binary_constraint5}, \eqref{eq:e>H}, \eqref{eq:h>abs} and \eqref{eq:h>-abs} so it is a linear programming task, which is computable in polynomial time \cite{KHACHIYAN198053}.

\subsubsection{Proof for directed graphs} \label{proof:directed LP}

The objective function for directed graphs can be written as
\begin{flalign}\label{eq:objective function directed}
    \underset{\substack{W\in\hW_{X,Y}},\,  \hat{e}\in \mathbb{R},\,  \hat{e}_2\in \mathbb{R}}{\arg\min}\left(\mathrm{tr}\big[D_{X,Y}^{\top}W\big]+\frac{\epsilon^p}{4} \hat{e} + \frac{\epsilon^p}{4} \hat{e}_2\right)
\end{flalign}
where we have introduced two extra variables $\hat{e}$, $\hat{e}_2 \in \R$. For  $\hat{e}$, the constraints are exactly the same as the case for undirected graphs, see \eqref{eq:constraint nonlinear}. For $\hat{e}_2$, the constraints are
\begin{flalign}
    \hat{e}_2 \geq \sum_{i=1}^{n_{X}}\sum_{j=1}^{n_{Y}}\left|\sum_{l=1}^{n_{Y}}A_{Y}(i,l)W(j,l) -\sum_{l=1}^{n_{X}}W(l,i)A_{X}(l,j)\right|,
\end{flalign}

To write the metric in LP form, we need to introduce the additional variables $H(i,j)$, see \eqref{eq:e>H}. In addition, similar to $H(i,j)$, we introduce additional variables $H_2(i,j) \in \R$ with constraints:
\begin{flalign}
    \hat{e}_2 &\geq \sum_{i=1}^{n_{X}}\sum_{j=1}^{n_{Y}} H_2(i,j), \label{eq:e2>H2}\\
    H_2(i,j) &\geq \sum_{l=1}^{n_{Y}}A_{Y}(i,l)W(j,l) -\sum_{l=1}^{n_{X}}W(l,i)A_{X}(l,j),  \label{eq:h2>abs}\\
    H_2(i,j) &\geq \sum_{l=1}^{n_{X}}W(l,i)A_{X}(l,j) -\sum_{l=1}^{n_{Y}}A_{Y}(i,l)W(j,l). \label{eq:h2>-abs}
\end{flalign}

Now, the metric computation can be written as an optimisation problem \eqref{eq:objective function directed}  with linear constraints \eqref{eq:binary_constraint1}, \eqref{eq:binary_constraint2}, \eqref{eq:binary_constraint3}, \eqref{eq:binary_constraint4}, \eqref{eq:binary_constraint5}, \eqref{eq:e>H}--\eqref{eq:h>-abs}, and \eqref{eq:e2>H2}--\eqref{eq:h2>-abs}, therefore, it is also computable in polynomial time.

\section{}\label{sec:proof of prop3: metric without attributes}

In this appendix, we prove Lemma \ref{lem:graph metric no attibutes}. If the base metric $d(\cdot,\cdot)$ is set to zero, from \eqref{eq:Dxy}, we have
\begin{align}
D_{X,Y} & =\left(\begin{array}{cc}
0_{n_{X},n_{Y}} & \frac{c^{p}}{2}1_{n_{X},1}\\
\frac{c^{p}}{2}1_{1,n_{Y}} & 0
\end{array}\right)\\
D_{X,Y}^{\top} & =\left(\begin{array}{cc}
0_{n_{Y},n_{X}} & \frac{c^{p}}{2}1_{n_{Y},1}\\
\frac{c^{p}}{2}1_{1,n_{X}} & 0
\end{array}\right).
\end{align}.
Then, we directly obtain
\begin{flalign}
\mathrm{tr}\left[D_{X,Y}^{\top}W\right] =\frac{c^{2}}{2}&\left(\sum_{i=1}^{n_{X}}W\left(i,n_{Y}+1\right) \right.
\nonumber \\
&+\left. \sum_{j=1}^{n_{Y}}W\left(n_{X}+1,j\right)\right)
\end{flalign}
This proves Eq. \eqref{eq:LP graph metric no attributes}. The symmetry property is direct to check. The triangle inequality proof in Appendix \ref{sec:proof of triangle inequality} is 
valid for $d\left(\cdot,\cdot\right)=0$. In addition, it is direct
to show that $\tilde{d}^{(c,\epsilon)}\left(X,X\right)=0$, which
finishes the proof that $\tilde{d}^{(c,\epsilon)}\left(\cdot,\cdot\right)$
is a pseudometric.

\end{document}